\documentclass[a4paper,12pt]{article}

\usepackage[dvips]{graphicx}
\usepackage{epsfig,amsmath,amssymb,verbatim,mathrsfs,array,layout,textcomp,amssymb,latexsym}
\usepackage{hyperref}

\newcommand{\beq}{\begin{eqnarray}}
\newcommand{\eeq}{\end{eqnarray}}
\newcommand{\nn}{\nonumber}

\newcommand{\idt}{\int d^2\theta}

\newcommand{\Z}{\mathcal{Z}}

\def\lsim{\mathrel{\rlap{\lower4pt\hbox{\hskip1pt$\sim$}}
     \raise1pt\hbox{$<$}}}         
\def\gsim{\mathrel{\rlap{\lower4pt\hbox{\hskip1pt$\sim$}}
     \raise1pt\hbox{$>$}}}         

\begin{document}

\begin{titlepage}

\begin{center}
\vskip1.5cm
{\huge \bf Vacuum (Meta)Stability \\Beyond the MSSM}\\
\end{center}
\vskip0.2cm

\begin{center}
{\large Kfir Blum, C\'edric Delaunay, Yonit Hochberg}
\end{center}
\vskip 1pt

\begin{center}
{\it Department of Particle Physics\\
Weizmann Institute of Science,\\ Rehovot 76100, Israel} \vspace*{0.3cm}

{\tt  kfir.blum@weizmann.ac.il,
cedric.delaunay@weizmann.ac.il,\\ yonit.hochberg@weizmann.ac.il}
\end{center}

\vglue 0.3truecm

\begin{abstract}
\vskip 3pt \noindent

We study the stability of the Higgs potential in the framework of the effective Lagrangian beyond the MSSM.
While the leading nonrenormalizable operators can shift the Higgs boson mass above the experimental bound, they
also tend to render the scalar potential unbounded from below. The destabilization is correlated with the Higgs
mass increase, so that if quantum corrections are small the problem is severe. We show that a supersymmetric
sub-leading correction stabilizes the potential within the domain of validity of the effective theory.
Constraints on MSSM parameters as well as on higher dimensional operators are derived, ensuring that our vacuum
has a lifetime longer than the present age of the universe. In addition we show that when effective operators
are responsible for evading the LEP bound, stability constraints imply an upper bound on the scale of new
physics in the few TeV range.

\end{abstract}

\end{titlepage}

\section{Introduction}

The Higgs tree level quartic couplings in the Minimal Supersymmetric Standard Model (MSSM) are completely
dictated by gauge interactions.  This fact stands at the heart of theoretical and phenomenological
difficulties, perhaps foremost among them the tree level prediction that the lightest Higgs boson be lighter
than the $Z$, a prediction ruled out by collider experiments \cite{Barate:2003sz, Schael:2006cr}. In order to
evade the experimental bound on the Higgs mass large quantum corrections are required in the MSSM, implying a
substantial hierarchy between the electroweak (EW) scale and the scale of supersymmetry (SUSY) breaking. In
particular, at least one of the top superpartners is required to be much heavier than the top such that some
amount of fine-tuning is needed~\cite{Chankowski:1997zh, Barbieri:1998uv, Chankowski:1998xv, Kane:1998im}.  The
fine-tuning becomes even more pronounced if the model is expected to provide explanations for present
cosmological data. While the MSSM possesses all of the ingredients needed in order to account for both the dark
matter and the baryon asymmetry of the universe, analyses reveal that difficulties associated with the Higgs
mass bound are rooted in the cosmological arcade, too (see, for example~\cite{ArkaniHamed:2006mb, Ellis:2007by,
Carena:2008rt}).

The restricted structure of the Higgs sector also makes it susceptible to small corrections from new physics
Beyond the MSSM (BMSSM). If the scale associated with the BMSSM physics lies well above MSSM particle masses,
an effective field theory approach becomes useful.  The effective theory framework of the BMSSM was previously
studied, \textit{e.g.}, in~\cite{Brignole:2003cm, Dine:2007xi, Antoniadis:2008es}.  The authors
of~\cite{Dine:2007xi} showed that the effective expansion takes on a rather simple form. In fact, under mild
assumptions the leading nonrenormalizable corrections to the Higgs sector are captured by only two operators,
one supersymmetric and the other associated with hard SUSY breaking (see also~\cite{Strumia:1999jm}). It was
further demonstrated that these operators may lead to a sizable shift of the Higgs mass at the classical level.
Besides potentially solving the difficulties mentioned above \cite{Casas:2003jx, Cassel:2009ps, Blum:2008ym},
this result is of considerable experimental importance as it opens up a zone of SUSY phenomenology in which
both stops may be very light, possibly just around the corner for collider experiments.

However, examining the vacuum structure of the effective theory reveals that when the leading nonrenormalizable
operators are taken to account for a significant shift for the lightest Higgs mass, they also destabilize the
quartic couplings.  Naively, this might be considered a severe setback to the picture drawn above.  In the
presence of a negative quartic coupling, stability relies on the higher order terms of the theory, which could
\textit{a priori} complicate the analysis considerably as well as introduce UV sensitivity.  Nevertheless, we
find that the effective theory exhibits a remarkable property: a single higher order operator arising at
dimension six automatically cures the runaway initiated by the leading dimension five terms.  Under mild
assumptions, this operator is the only one relevant to the question of stability, and, furthermore, is directly
correlated with the dimension five set.  We find that the entire study of potential stability can be conducted
and resolved within the range of validity of the effective approach. As a result, vacuum stability can be
ensured by imposing simple relations between MSSM parameters and dimension five BMSSM operators, without the
need to explicitly invoke the potentially far more complicated dimension six structure.  In addition, we show
that an upper bound on the scale of new physics may arise in the BMSSM, as well as derive the viable range for
the lightest Higgs mass.

The interplay between the negative quartic and the higher dimensional operator often leads to the formation of
a remote vacuum, in the presence of which the EW vacuum is only metastable.  The existence of the remote vacuum
was noticed by the authors of~\cite{Batra:2008rc}, where it was suggested that EW symmetry breaking may in fact
lead our universe to reside in this new configuration dominated by nonrenormalizable operators.  An intriguing
feature of this scenario is that in this case, EW breaking may occur even in the SUSY limit. However, the
structure of the effective theory implies that the EW scale be given in this case by the geometric mean of the
new physics scale $M$ and the SUSY $\mu$ parameter\footnote{The emergence of an intermediate scale of similar
formal form was previously considered in~\cite{Martin:1999hc}, where it was discussed in the context of much
higher energy phenomena.}, $v\sim\sqrt{\mu M}$. Null searches for charginos, for instance, highly restrict this
possibility. In this paper we adopt a more conservative approach. We discuss the possibility that while the
leading effective operators act to destabilize the potential, the usual EW breaking vacuum is either stable or
of a life time longer than the present age of the universe.

The outline of this paper is as follows.  In Section \ref{sec:BMSSM} we review the BMSSM Higgs
sector, emphasizing the correlation between the lightest Higgs mass and the appearance of
directions in field space where the potential is unbounded from below at leading order. We then study the next set of higher
dimensional operators, showing that they can stabilize the potential up to the cutoff scale of the effective
theory.  In Section \ref{sec:vacstab} we present analytical and numerical prescriptions ensuring the
(meta)stability of the EW vacuum.  Section \ref{sec:phen} discusses the phenomenological implications of the
stability constraints.  Our conclusions are gathered in Section \ref{sec:disc}.  Appendix
\ref{app:AppPotential} addresses the issue of charge breaking and CP violating field configurations, Appendix
\ref{app:qtunn} elaborates on the quantum tunneling computation, and Appendix \ref{app:EWPT} presents
constraints on the heavy cutoff scale arising from electroweak precision measurements.

\section{Higgs sector in the MSSM and Beyond}\label{sec:BMSSM}

In the bulk of this paper we analyze the vacuum stability of the BMSSM theory, namely the MSSM Higgs sector
augmented by nonrenormalizable operators. It is useful to adopt the nomenclature of \cite{Dine:2007xi} and to
classify the nonrenormalizable operators according to their {\it scaling} dimension, which is the total mass
dimension of the fields contained in an operator, and their {\it effective} dimension, counting the powers of
$1/M$ which suppress it.
For instance, an operator such as $(\mu/M)h_uh_d|h_u|^2+{\rm h.c.}$ is of scaling dimension four but effective
dimension five.

In Section~\ref{sec:mssm} we briefly discuss the renormalizable MSSM Higgs sector, and highlight the main
features which guide us in the study of nonrenormalizable corrections to it. In Section~\ref{sec:d5} we review
BMSSM operators of effective dimension five. These operators alone suffice to lift the lightest Higgs mass
above the LEP bound. However, as we show, these operators can destabilize the scalar potential at large field
values which are still within the domain of validity of the effective theory. In that case, naively, the
effective theory truncated at this order is not consistent; higher order operators must cure the instability.
The next set of operators consists of effective dimension six, and we study it in Section \ref{sec:d6}.

\subsection{MSSM setup}\label{sec:mssm}

The scalar Higgs potential of the renormalizable MSSM can be written as follows\footnote{We write the
superfield components as $H=(h,\psi,F)$, the MSSM $\mu$ term as $\int d^2\theta\mu H_uH_d$, and our convention
for SU(2) contraction is $H_uH_d=H_u^+H_d^--H_u^0H_d^0$.}:
\beq\label{eqn:scaltreeMSSM}
V_{\rm MSSM}&=&
m^2_1\left|h_d\right|^2+m^2_2\left|h_u\right|^2+\left(m^2_{12}h_uh_d+{\rm h.c.}\right)\nonumber\\ &&+
\frac{g_Z^2}{8}\left(\left|h_u\right|^2-\left|h_d\right|^2\right)^2+
\frac{g^2}{2}\left(|h_u|^2|h_d|^2-|h_uh_d|^2\right),
\eeq
where $m_1^2\equiv m^2_{H_d}+|\mu|^2$, $m_2^2\equiv m^2_{H_u}+|\mu|^2$, $m_{12}^2\equiv B\mu$ and
$g_Z^2=g^2+g'^2$ with $g, g'$ the SM gauge couplings. One can always choose a basis for the fields such that
$m_{12}^2$ is real and positive, and we shall keep to such a basis consistently throughout the paper. We
parameterize the expectation values as follows:
\beq\label{eq:VEVpar}\langle h_u\rangle^T=(0,\phi_2), \ \
\langle h_d\rangle^T=(\phi_1+i\chi,\rho).
\eeq
Gauge freedom is used to render $\langle h_u^+\rangle=0$ and $\phi_2$ real and positive, while the real
component $\phi_1$ is allowed to obtain negative values.\footnote{Note that, in the renormalizable MSSM, the
minimization conditions lead to $\rho=\chi=0$. In that case one may write the scalar potential using
non-negative $\phi_1$ and $\phi_2$.} In the BMSSM, as we shall see, more than one vacuum configuration may
develop. We will analyze paths of minimum potential energy which connect these vacua. Along such paths, in
principle, it may become energetically favorable for non-vanishing values of $\chi$ or $\rho$ to turn on. This
may occur even when all Lagrangian parameters are real, as we shall assume in this paper. In this case, we have
found that ignoring the CP violating (CPV) and charge-breaking (CB) background fields $\chi$ and $\rho$ is
typically well justified in the study of potential stability. Therefore, for clarity, we set $\rho=\chi=0$ in
most of the paper. In places where deviations from this assumption become relevant we refer the reader to the
discussion in Appendix \ref{app:AppPotential}.

In the framework we study, $\mathrm{SU(2)_L\times U(1)_Y}$ breaking into $\mathrm{U(1)_{EM}}$ occurs as usual
by an interplay between the quadratic and quartic terms in the scalar potential. (For an alternative scenario,
see~\cite{Batra:2008rc}.) The resulting vacuum is parameterized by
\beq\label{eq:vtb} &v^2=\phi_1^2+\phi_2^2\simeq (174 \ \mathrm{GeV})^2,\nn\\
&\tan\beta=\phi_2/\phi_1.\label{eq:VEV} \eeq
We call this vacuum the EW vacuum.

Being exclusively dictated by gauge superfield D-terms, the quartic couplings in (\ref{eqn:scaltreeMSSM}) are
proportional to the electroweak gauge couplings, making the Higgs sector sensitive to small corrections. In
particular, quantum corrections arising from loops of MSSM fields with a large amount of SUSY breaking are
usually conceived to account for a sizable shift in the lightest Higgs mass. In a similar manner, the quartic
Higgs structure is also sensitive to nonrenormalizable operators, which need only compete with couplings of
order $g^2$ in order to modify the spectrum.

Moreover, the quartic terms of \eqref{eqn:scaltreeMSSM} vanish along the D-flat directions, specified by
$|\langle h_u\rangle|=|\langle h_d\rangle|$. At tree (and renormalizable) level, this gives rise to a
constraint on the quadratic terms, $m_1^2+m_2^2>2m_{12}^2$, which must hold for the potential to be bounded
from below. When nonrenormalizable operators are considered, the relative flatness of the renormalizable
potential along the D-flat directions makes it important to verify that the higher dimensional operators do not
destabilize the vacuum.

\subsection{Operators of effective dimension five}\label{sec:d5}

Effective dimension five operators composed purely of Higgs fields enter the Lagrangian via the superpotential.
Including F-term SUSY breaking, there are two such operators~\cite{Dine:2007xi}:
\beq
\frac{1}{M}\idt\left(\lambda_1(H_uH_d)^2+\lambda_2\Z(H_uH_d)^2\right)+{\rm
h.c.}\label{eqn:dim5} \eeq
Here $\Z\equiv m_S\,\theta^2$ is a dimensionless chiral superfield spurion. One could also contemplate the
existence of operators arising from D-term SUSY breaking, in which case additional effective dimension five
operators arise. In this paper, however, we assume that the effect of D-term breaking is somehow suppressed or
non-existent~\cite{Dine:2007xi,Komargodski:2009pc}.

We assume that the new physics generating the effective operators is approximately supersymmetric, $m_S\ll M$.
We are interested in the imprint which the effective dimension five operators have on both the spectrum and the
stability of the scalar potential, as we now discuss. At effective dimension five, the correction to the scalar
potential resulting from (\ref{eqn:dim5}) reads:
\beq \delta V_5 =
2\epsilon_1h_uh_d\left(|h_u|^2+|h_d|^2\right)+\epsilon_2(h_uh_d)^2+{\rm h.c.},\label{eqn:dV5} \eeq where we
have defined $\epsilon_1\equiv \lambda_1\mu^*/M$ and $\epsilon_2\equiv-\lambda_2 m_S/M$. Expanding to order
$\mathcal{O}(\epsilon)$, the following shift is obtained for the light (CP-even) Higgs boson mass:
\beq\label{eqn:massDST}\delta_{\epsilon} m_h^2&=
&2v^2\left(\epsilon_{2r}-2\epsilon_{1r}\sin2\beta-
\frac{2\epsilon_{1r}(m_A^2+m_Z^2)\sin2\beta+\epsilon_{2r}(m_A^2-m_Z^2)\cos^22\beta}{\sqrt{(m_A^2-m_Z^2)^2+
4m_A^2m_Z^2\sin^22\beta}}\right),\eeq
where $\epsilon_{kr}$ denotes the real part of $\epsilon_k$. As explained in \cite{Dine:2007xi}, only the real
parts of $\epsilon_{1,2}$ enter the spectrum at leading order while the imaginary parts contribute to interactions
and mixing. In this paper we make use of the fact that the spectrum is relatively insensitive to the imaginary parts
of $\epsilon_{1,2}$ and consider, for simplicity, only the case where there exists a basis for $h_u,h_d$ in which
$\epsilon_{1,2}$ are real and $m_{12}^2$ is real and positive. Henceforth we drop the $r$ subscript and assume that
$\epsilon_{1,2}$ are real.

\begin{figure}
\includegraphics[width=8cm]{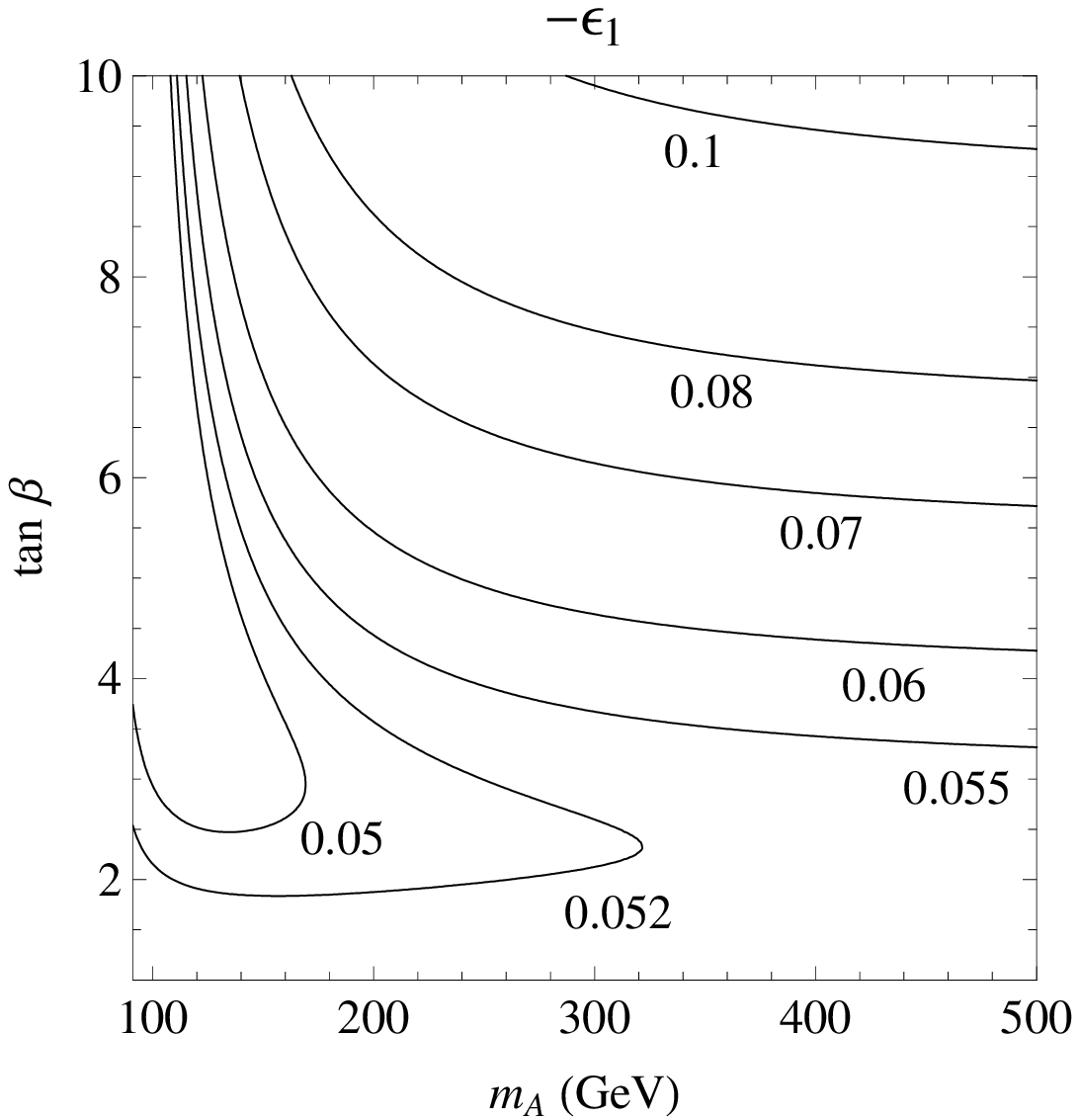}\hfill
\includegraphics[width=8cm]{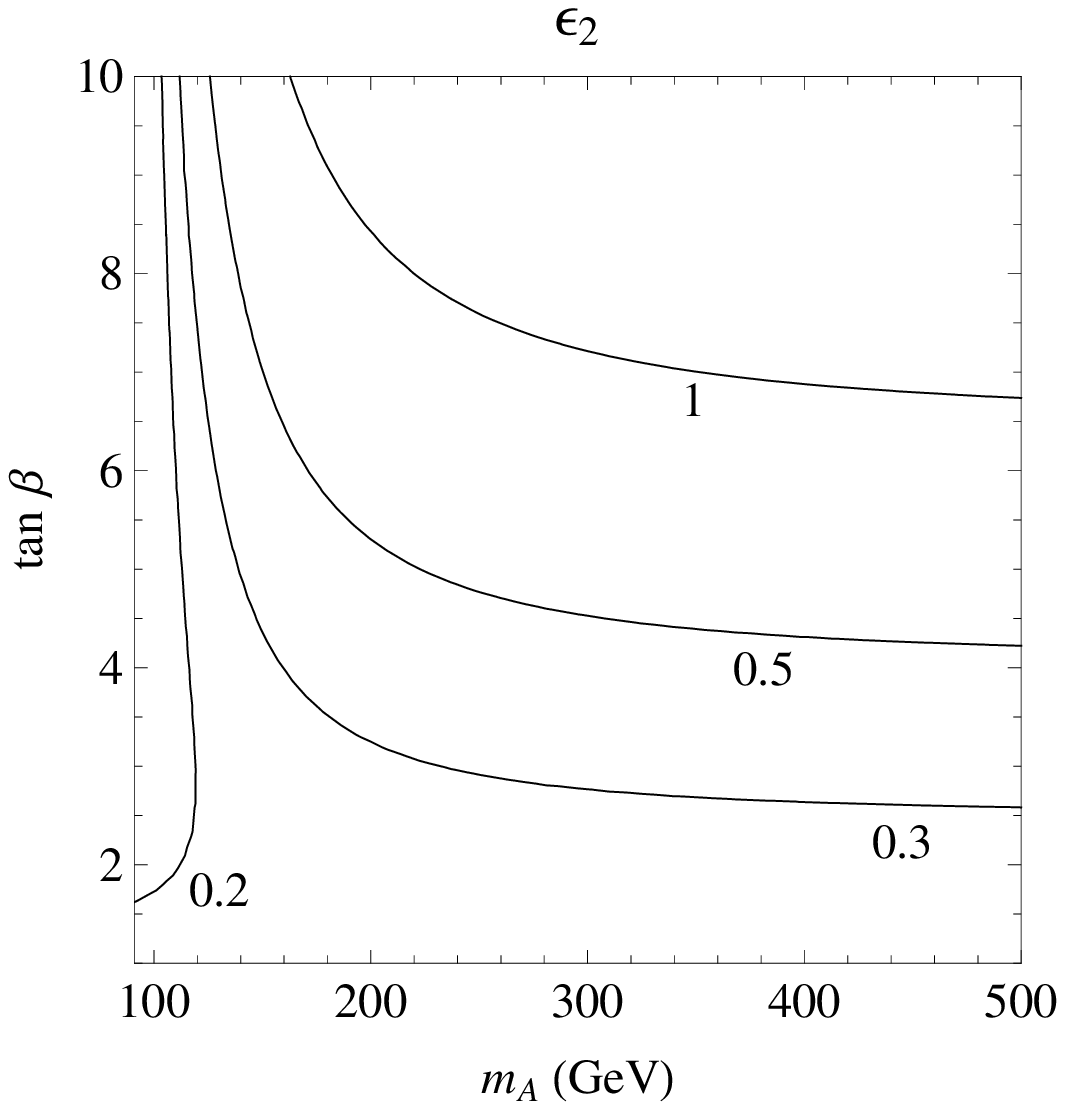}
\caption{Contour plots of the values of $-\epsilon_1$ (left) and $\epsilon_2$ (right) corresponding to a fixed
Higgs mass $m_h=115$ GeV. In each panel only the considered operator is nonzero.}\label{fig:epscontours}
\end{figure}

The contribution of the non-supersymmetric term to the mass shift is suppressed compared to the supersymmetric
one. For example, in the limit $m_A^2\gg m_Z^2$ we have
\beq\label{eq:mhLargemA}
\frac{m_h^2}{m_Z^2}= \cos^22\beta+\frac{4\epsilon_2\sin^22\beta}{g_Z^2}-\frac{16\epsilon_1\sin2\beta}{g_Z^2}+
\mathcal{O}\left(\frac{m_Z^2}{m_A^2}\right).
\eeq
It follows that $|\delta_{\epsilon_2} m_h^2/\delta_{\epsilon_1}
m_h^2|\approx|\epsilon_{2}/\epsilon_{1}|\sin2\beta/4$, and so $\epsilon_{2}>4|\epsilon_{1}|$ is needed in order
for both terms to give a comparable mass shift. In Figure~\ref{fig:epscontours} we report the size of both the
SUSY preserving and breaking operators, as required for $m_h=115$ GeV, illustrating that the former can easily
lift the Higgs mass classically at this order, while existence of only the latter calls for large quantum
corrections as in the MSSM.
For moderate $\tan\beta$ a negative $\epsilon_1$ of magnitude
$|\epsilon_{1}|\gtrsim10^{-2}\times(1-r^{-2})\tan\beta$, with $r\equiv m_A/m_Z$, can lift the Higgs mass above
the LEP bound at tree level, without the need for quantum corrections. The large $\tan\beta$ limit is more
involved. For $\tan\beta>|1/\epsilon_1|$, the leading contribution of effective dimension five operators is
suppressed to the level of the next order in $\epsilon$ and, as a result, becomes comparable to that of
effective dimension six terms.

Apart from affecting the local properties of the vacuum ({\it e.g.} particle spectrum and interactions) the
effective dimension five operators also influence the potential at large field values. Along the D-flat
directions\footnote{Note that there are now two such directions as in our convention $\phi_1$ can be negative.}
the potential is only bounded from below as long as $4|\epsilon_1|<\epsilon_2$. If this relation does not hold,
the potential eventually becomes unstable and a runaway occurs along the direction $\phi_1={\rm
sign}(\epsilon_1)\phi_2$. In the interesting case where $\epsilon_1<0$, corresponding to a positive shift to
the lightest Higgs mass, a saddle point emerges at field values \beq\label{eq:saddle5}
-\phi_1=\phi_2=\sqrt{\frac{m_1^2+m_2^2+2m_{12}^2}{8\left(|\epsilon_{1}|-\frac{1}{4}\epsilon_2\right)}},
\eeq
after which the potential decreases indefinitely. An important property of the runaway is that it occurs at
field values $\phi^2\sim m^2/\epsilon\sim m M$, where $m$ refers to a combination of relevant quadratic mass
parameters (presumably of electroweak scale) and $M$ is the heavy scale. Therefore, the runaway develops at
$\phi\ll M$, rendering the EW vacuum unacceptably short-lived.

Given the discussion above, one might naively conclude that the inequality $4|\epsilon_1|<\epsilon_2$ must be
enforced as a physical constraint on the relative magnitudes of the supersymmetric vs. the SUSY breaking
corrections~\cite{Antoniadis:2008es}. In particular, in the supersymmetric limit wherein $\epsilon_2=0$, also
$\epsilon_1=0$ would be required. However, as we show below, the supposed runaway is an artifact of the
truncation of the effective expansion at $\mathcal{O}(M^{-1})$~\cite{Casas:2001xv}. This expansion cannot be
trusted at large field values, even though it is consistent in the local vicinity of our
vacuum~\cite{Batra:2008rc}. Indeed, effective dimension six operators generate positive scaling dimension six
contributions of the form $\phi^6/M^2$. Such terms eventually win over the negative quartic terms at field
values $\phi^2\sim m M$, precisely the region of field space at which the instability starts to develop. Thus,
effective dimension six operators are {\it as important as} the effective dimension five ones around the
instability and may stabilize the potential before the UV threshold of the effective theory. Below we inspect
the structure of effective dimension six operators in order to determine which ones are relevant to the
stability problem. Compared to previous studies~\cite{Piriz:1997id,Cassel:2009ps}, we find that focusing on the
physical problem at hand simplifies the analysis considerably. Delightfully, we are able to show that over a
broad range of parameters only one such operator exists, and even this operator not strictly independent from
the dimension five set.

\subsection{Operators of effective dimension six}\label{sec:d6}

We begin by considering effective dimension six operators which involve SUSY breaking. At order $1/M^2$, there
are no scaling dimension six operators which arise from SUSY breaking terms in the Lagrangian. The reason is
that each term arising from SUSY breaking spurion must be accompanied by the appropriate power of the SUSY
breaking mass scale. At scaling dimension six, this implies that the first (pure Higgs) operators are of order
$1/M^3$.

We do find effective operators associated with SUSY breaking spurions at scaling dimension four. These are of
the form $(m/M)^2h^4$, where we extend the definition of $m$ to include the SUSY breaking scale, also of order
the electroweak scale in this framework. Along the D-flat directions such contributions are suppressed by an
additional power of $m/M$ in comparison to the effective dimension five terms of $\delta V_5$. Away from the
D-flat directions, MSSM D-terms guarantee stability provided that we impose
\beq\label{eq:sc4eff6} \epsilon^2\lsim\frac{g_Z^2}{8}\approx\frac{1}{15}\eeq
as a parametric inequality, where $\epsilon^2\equiv(m/M)^2$ represents the magnitude of the dimensionless
coefficient of such operators. Henceforth we restrict the discussion to the scenario in which
Eq.~(\ref{eq:sc4eff6}) is satisfied. We shall see that this assumption simplifies the problem considerably.
Moreover, it stands in accordance with requiring the effective theory expansion in powers of $\epsilon$ to
remain valid, regardless of the stability analysis. Referring to Figure~\ref{fig:epscontours} we find that, at
least for moderate $\tan\beta$, Eq.~(\ref{eq:sc4eff6}) does not pose any real limitation on the role of the
$\epsilon_1$ correction in lifting the Higgs mass above the current experimental bound. Nevertheless, in
general, the scaling dimension four, effective dimension six operators do not display any particular
$\tan\beta$ dependence and so they can have a comparable influence on the spectrum in the large $\tan\beta$
regime, where the leading dimension five contributions are suppressed. Due to the multiplicity of independent
coefficients, the analysis of the spectrum is involved in this case and we do not pursue it further in this
paper.

From the discussion above we conclude that it is enough for our purpose to study effective dimension six
operators in the supersymmetric limit. Squaring the $1/M$ piece of the F-term equations of motion, arising in
the presence of the supersymmetric effective dimension five operator gives
\beq\label{eqn:dV6} \delta V_6=
4\left|\frac{\epsilon_1}{\mu}\right|^2|h_uh_d|^2\left(|h_u|^2+|h_d|^2\right).
\eeq
The contribution (\ref{eqn:dV6}) is positive definite and non-vanishing along the D-flat directions. Thus it
plays an important role in stabilizing the potential. Note that, for a given value of the $\mu$ parameter, the
coefficient of this contribution is correlated with the Higgs mass through Eq.~(\ref{eqn:massDST}).

Superpotential operators which involve gauge superfields contribute to the scalar potential at effective
dimension six, and must contain D-term components. $\mathrm{K\ddot{a}hler}$ operators which arise at order
$1/M^2$ can only affect the scalar potential through either F-terms or D-terms, by gauge invariance.
Considering F-terms, we immediately see that these cannot contribute to effective dimension six, scaling
dimension six operators, since they are linear in $h$ at leading order in $1/M$. F-terms do contribute to
effective dimension six, scaling dimension four operators. However, following Eq.~(\ref{eq:sc4eff6}) and the
related discussion, such contributions are parameterically suppressed and we need not pursue them further.
Considering D-terms, and ignoring sub-leading scaling dimension four operators, we find scaling dimension six
operators which are of the general form
\beq \frac{\tilde g^2}{M^2}h^4\left(|h_u|^2-|h_d|^2\right) \eeq
where $\tilde g^2$ stands for some bilinear combination of $g$ and $g'$.

We learn that, apart from the operator (\ref{eqn:dV6}), all other scaling dimension six contributions to the
scalar potential are gauge coupling suppressed and -- most importantly -- vanish along the D-flat directions of
the MSSM. Therefore at large field values (but still $\ll M$) and along the would-be runaway direction
described in the previous section, the potential is in fact driven by the positive definite scaling dimension
six $\delta V_6$ given in (\ref{eqn:dV6}). Thus the same superpotential operator responsible for lifting the
Higgs mass classically also ensures that the scalar potential does not exhibit a runaway. The phenomenology of
this correlation we aim to address in the next section.

\section{Vacuum stability}\label{sec:vacstab}

In what follows we analyze the stability of the scalar potential including the dimension six operator
identified in the previous section, namely $V=V_{\rm MSSM}+\delta V_5+\delta V_6$. Simple analytical and
numerical criteria which ensure the stability of a given potential configuration are formulated. These criteria
can be put in terms of relations between the electroweak scale parameters and effective dimension five
operators.

\subsection{Analytical approximation}\label{ssec:anal}

It is useful to first analyze the potential along the MSSM D-flat directions. While soft terms, quantum
corrections and the presence of the EW vacuum all play a role in shifting the potential features somewhat away
from $|\langle h_u\rangle|=|\langle h_d\rangle|$, all of the insights are contained and, further more, it turns
out to be a reasonable approximation to study the profile of the potential at these well defined directions in
field space. We begin by performing this analysis at tree level, assuming vanishing $\chi$ and $\rho$ values.
Then, having obtained the principal results we extend the discussion to include all of the complications
mentioned above.

At tree level, the effective potential along the MSSM D-flat directions take the form:
\beq\label{eq:Vflat} V^{\rm D-flat}(\phi)=
\frac{1}{2}\left(m_2^2+m_1^2\mp 2m_{12}^2\right)\phi^2+2\left(\frac{\epsilon_2}{4}\mp \epsilon_1\right)\phi^4+
\left|\frac{\epsilon_1}{\mu}\right|^2\phi^6. \eeq
with $\pm\phi_1=\phi_2\equiv\phi/\sqrt{2}$. If the quartic coupling in (\ref{eq:Vflat}) is negative, the
potential may develop another vacuum away from the electroweak scale. Thus a simple way to guarantee stability
is to impose $\epsilon_2>4|\epsilon_1|$. However, from Figure \ref{fig:epscontours} we learn that typical
values for $|\epsilon_1|$, consistent with the bound on the lightest Higgs mass are in the range
$|\epsilon_1|\sim0.05-0.1$. The simple condition for an always-positive quartic is therefore in some tension
with the need to assure that higher order terms are under control (see Eq.~(\ref{eq:sc4eff6})). Hence we attend
to the more interesting case where the quartic coupling is negative. It is useful to define the quantities
$m^2\equiv m_2^2+m_1^2\mp 2m_{12}^2$ and $\tilde\epsilon\equiv\epsilon_2/4\mp \epsilon_1$. Using these
quantities we see that the non-zero extrema of the potential, corresponding to the remote vacuum and saddle
point are located respectively at
\beq\label{eq:remvac}\phi^2=-\frac{2|\mu|^2}{3\tilde{\epsilon}}\left(\frac{\tilde{\epsilon}}{\epsilon_1}\right)^2
\left[1\pm\sqrt{1-\frac{3 m^2}{8|\mu|^2}\left(\frac{\epsilon_1}{\tilde{\epsilon}}\right)^2}\right].\eeq
Manipulating Eqs.~(\ref{eq:Vflat}) and (\ref{eq:remvac}) we arrive at the following criterion,
designed to ensure that the remote minimum along the D-flat direction is at most degenerate with, but never
deeper than the potential at the origin of field space:
\beq\label{eq:vaccond}
\frac{ m^2}{|\mu|^2}\geq2\left(\frac{\tilde{\epsilon}}{\epsilon_1}\right)^2.
\eeq

A remarkable feature of the condition (\ref{eq:vaccond}) is that it does not involve the heavy scale $M$.
Rather, it involves a relation between relevant mass parameters of the MSSM and the relative size of
nonrenormalizable operators. Indeed, from Eq.~(\ref{eq:remvac}) it follows that both the remote vacuum and the
saddle point scale similarly, as $\phi\sim\sqrt{mM}\ll M$. This behavior reflects the fact that the quartic
couplings along the D-flat directions originate from effective dimension five operators. As a result, there
exists a scale at which quadratic, quartic and dimension six contributions in the potential are of similar
magnitude; this scale is precisely $\sqrt{mM}$.

For concreteness we attend to the case $\epsilon_1<0$, corresponding to a positive shift for the lightest Higgs
mass. Rewriting Eq.~(\ref{eq:vaccond}) in terms of physical quantities and to leading order in $\epsilon$, we
arrive at the following useful relation:
\beq\label{eq:VstabEW}
\frac{m_A^2(1+\sin2\beta)}{|\mu|^2}\geq
2\left(\frac{\tilde\epsilon}{\epsilon_1}\right)^2
\left[1+\frac{m_Z^2}{m_A^2}\frac{16\tilde\epsilon}{g_Z^2}
\left(\frac{1+2\sin2\beta}{1+\sin2\beta}-\frac{3}{2}\frac{\epsilon_1}{\tilde\epsilon}\right)\right]^{-1}.
\eeq
Restricting to the D-flat direction allowed us to write a stability criterion of simple analytical form.
However, the actual vacua and saddle point emerge somewhat away from the D-flat direction. In particular, the
remote vacuum (and, similarly, saddle point) can sustain an angle $\sim m/\sqrt{mM}\sim\sqrt{\epsilon}$ from
the flat direction. Hence imposing stability along $|\phi_1|=|\phi_2|$ does not forbid the actual remote vacuum
from becoming a global minimum. It is therefore important to complement Eq.~(\ref{eq:VstabEW}) using numerical
methods. In practice, as we show below, Eq.~(\ref{eq:VstabEW}) turns out to be a robust but slightly
conservative stability criterion.

\subsection{Numerical approach}\label{ssec:numer}

Armed with intuition from the analytical analysis, we proceed to define stability criteria based on
numerical procedures. We formulate two such criteria, then comment on the possibility of CPV or CB field
configurations.

\paragraph{Vacuum degeneracy}
First, we define a stability criterion by computing the potential and numerically verifying that the remote
vacuum is at most degenerate with, but never deeper than the EW vacuum. We call this constraint vacuum
degeneracy; it is robust but conservative.

\paragraph{Quantum tunneling}
Second, we define a stability criterion by numerically computing the tunneling rate from the EW to the remote
vacuum. The tunneling rate is given by $\Gamma\propto\exp\left(-B\right)$, with $B$ the bounce action. A
metastable configuration is viable if the universe remains in the EW vacuum for longer than its age.
Quantitatively, this translates into $B\gsim400$. We have used an approximate method to compute the bounce
action, and errors of $\mathcal{O}(1)$ are expected. Hence we discuss configurations with bounce action of
$B=400$ and $10^3$. A detailed computation of the tunneling rate can be found in Appendix \ref{app:qtunn}.

\paragraph{}
In the discussion above we have ignored the CP violating and charge breaking background fields, $\chi$ and
$\rho$. However, these effects are accounted for when we numerically search for vacuum degeneracy and compute
the bounce action. We find that neglecting the $\chi$ and $\rho$ background fields is almost always justified
in the analysis of potential stability in our framework, and does not affect the reliability of
Eq.~(\ref{eq:VstabEW}). The reason is that the stability criterion itself is typically sufficient to ban
non-vanishing $\chi$ and $\rho$, for field values $\phi<M$. More details can be found in Appendix
\ref{app:AppPotential}.

In Figure \ref{fig:stabcritM} (three panels on the left) the stability criteria are projected on various
sections of the parameter space, illustrating the above analysis. We focus on the combined study of electroweak
scale parameters vs. the heavy scale $M$. The following statements can be made.

The analytical stability criterion is slightly less conservative than, but closely follows the robust numerical
criterion of vacuum degeneracy. Moreover, in the parameter space depicted in Figure \ref{fig:stabcritM} we find
that the analytical criterion is typically more conservative than imposing $B\gsim10^3$, and always ensures
that the bounce action will be above $B\sim400$. Thus we can safely say that the analytical criterion is
robust, at least for moderate values of $\tan\beta\lsim10$ where small $\epsilon_1\lsim0.1$ is more than sufficient to
lift the Higgs mass above the experimental bound.
Since the gap between the various criteria is modest at all parameter values, it is evident that
Eq.~(\ref{eq:VstabEW}) provides a detailed qualitative, as well as quantitative understanding of the parameter
space of vacuum stability.

Regarding the tunneling action, we find that it rises steeply above $B\sim100$, rendering the $B=400$ and
$10^3$ contours very close to each other. This occurs since in the relevant regions of parameter space the
remote minimum is nearly degenerate with the EW vacuum, and so a small change of parameters towards vacuum
degeneracy causes the action to diverge.

\begin{figure}
\includegraphics[width=8.5cm]{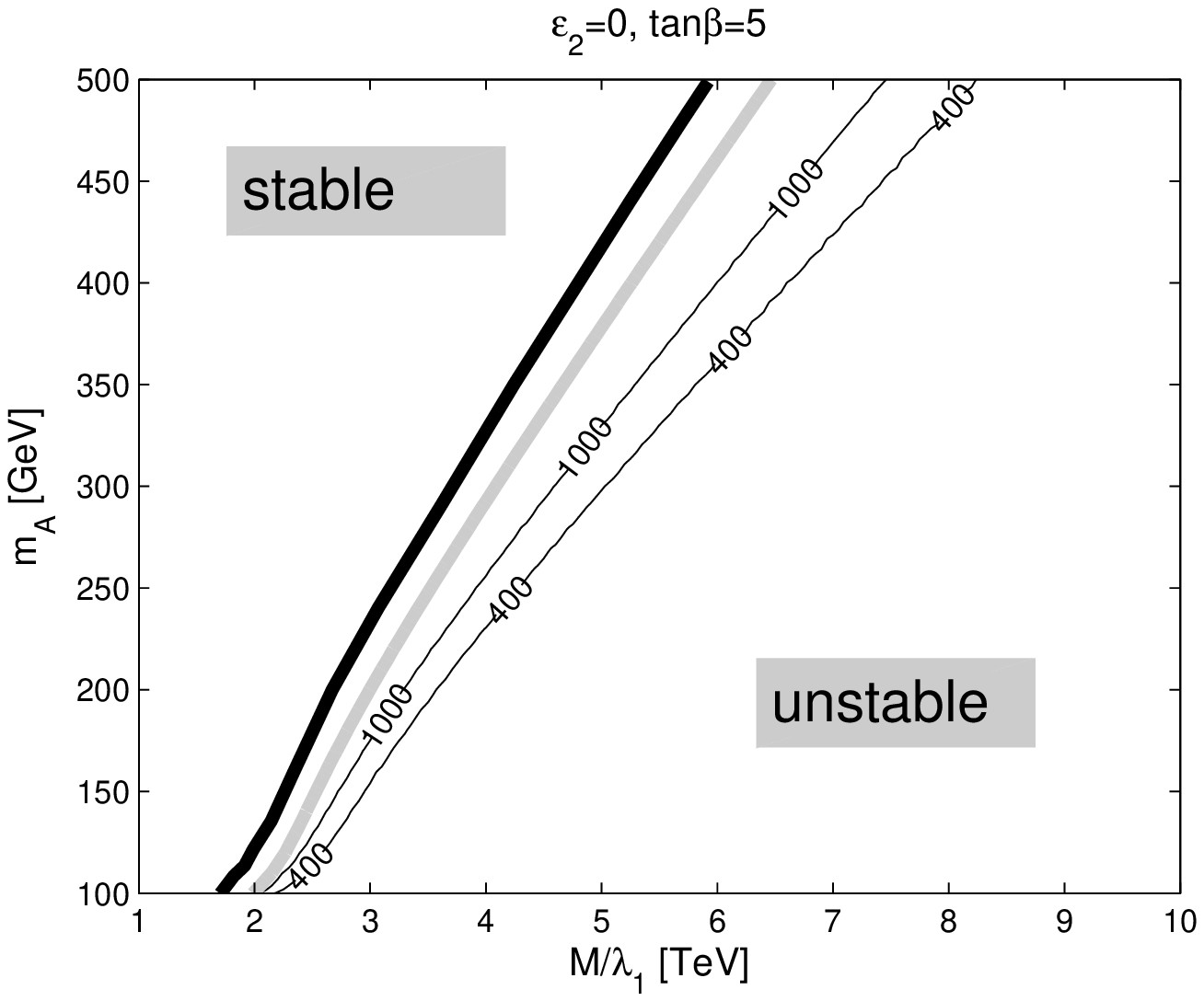}\hfill
\includegraphics[width=8.5cm]{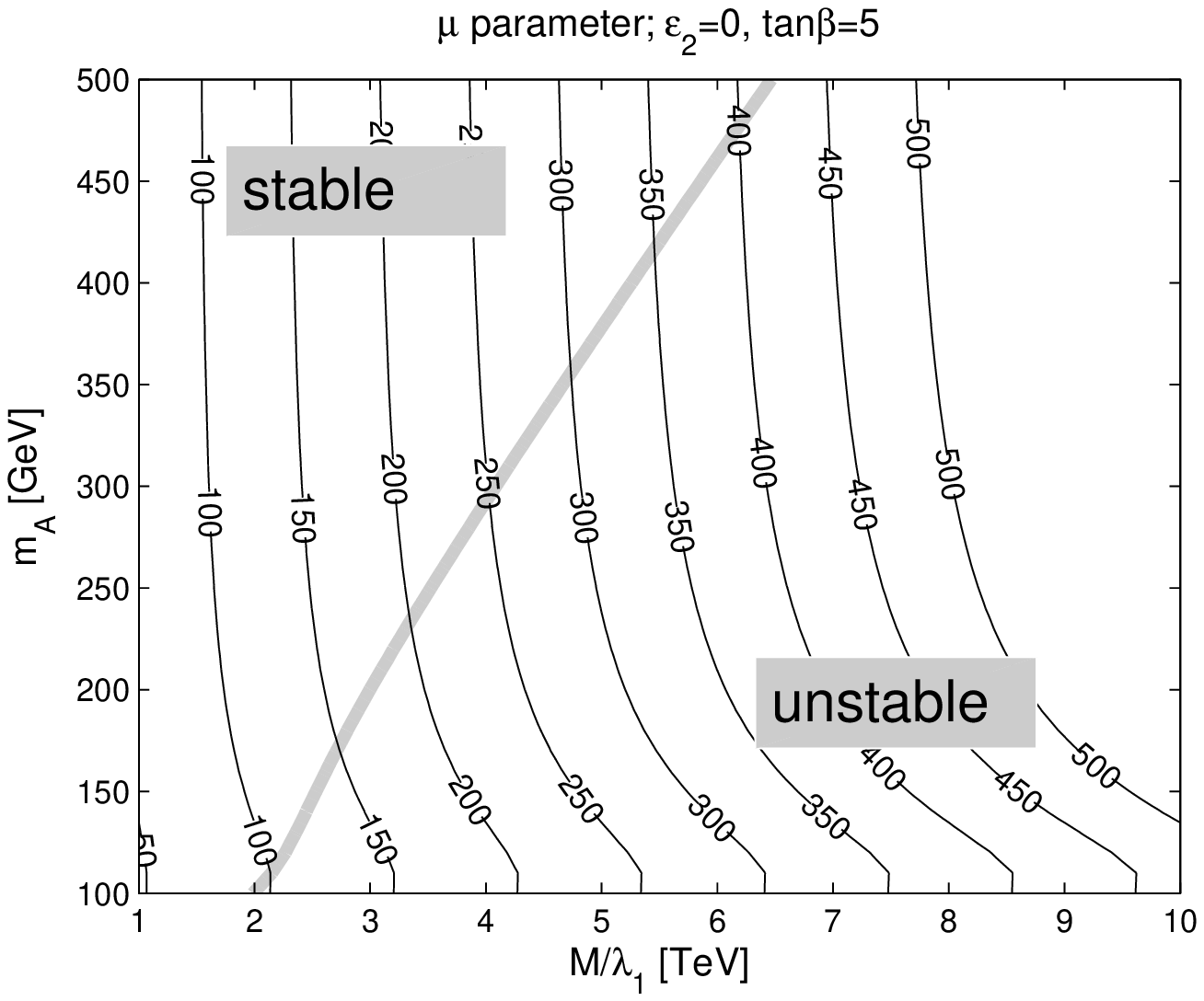}\\
\includegraphics[width=8.5cm]{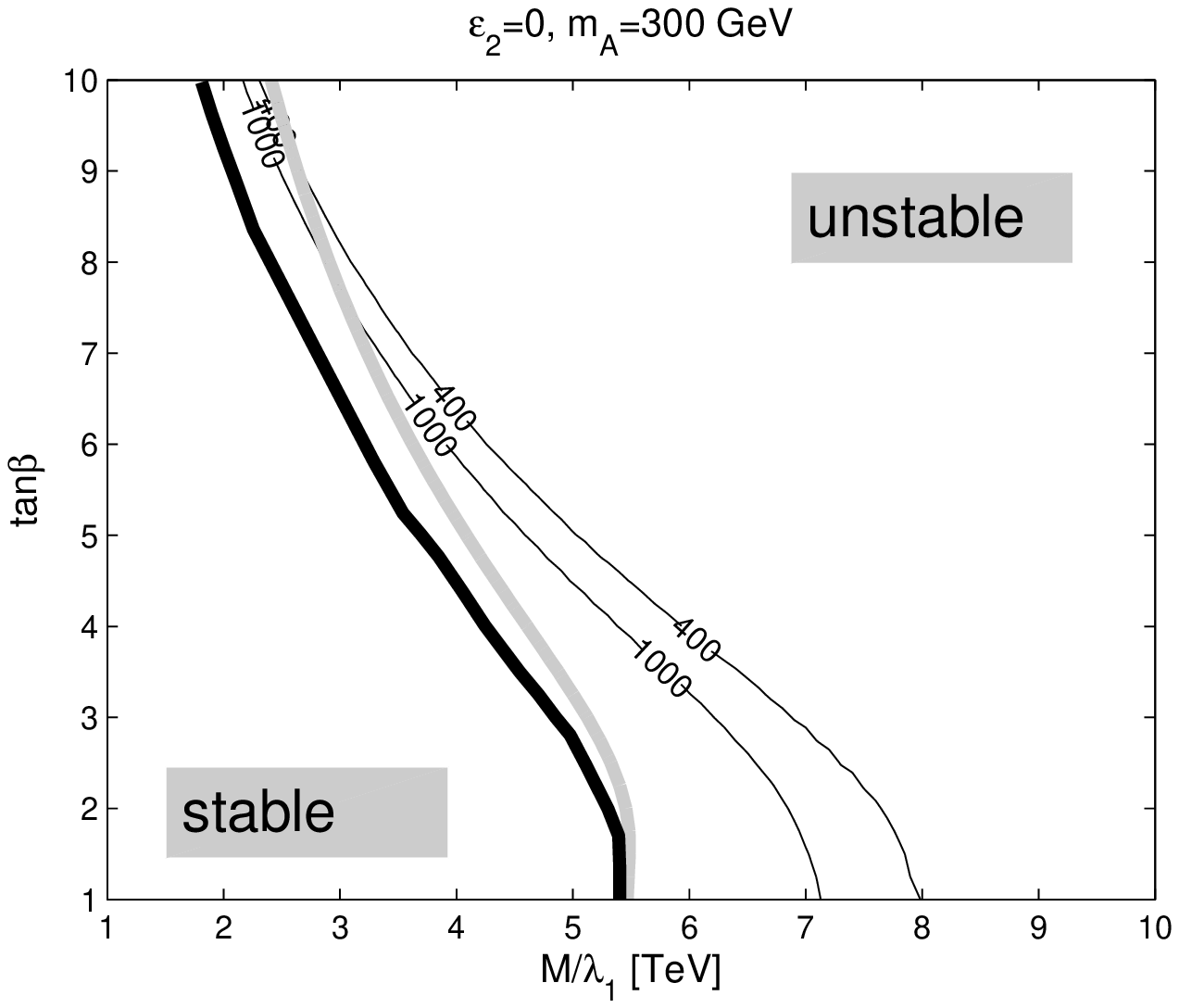}\hfill
\includegraphics[width=8.5cm]{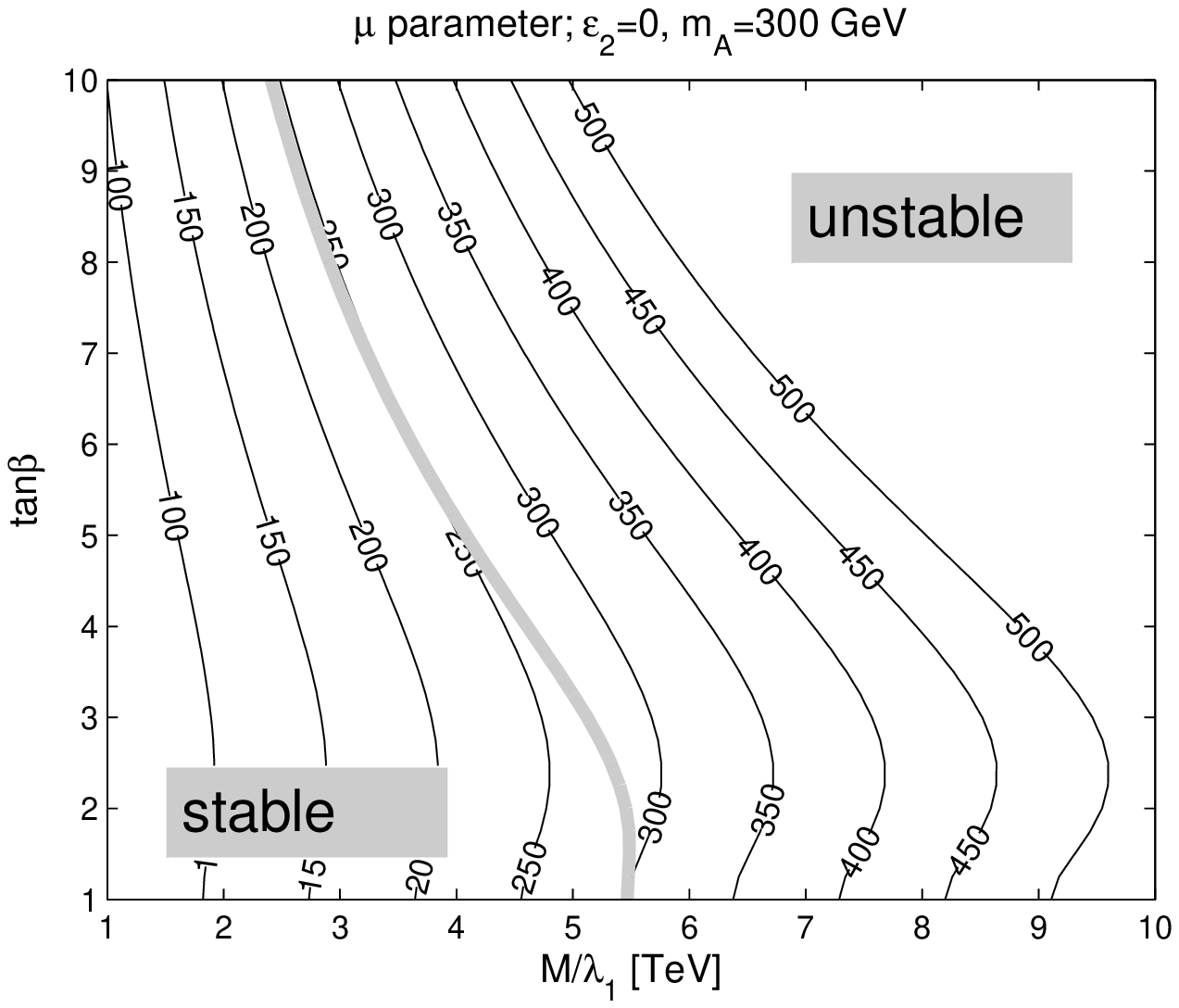}\\
\includegraphics[width=8.5cm]{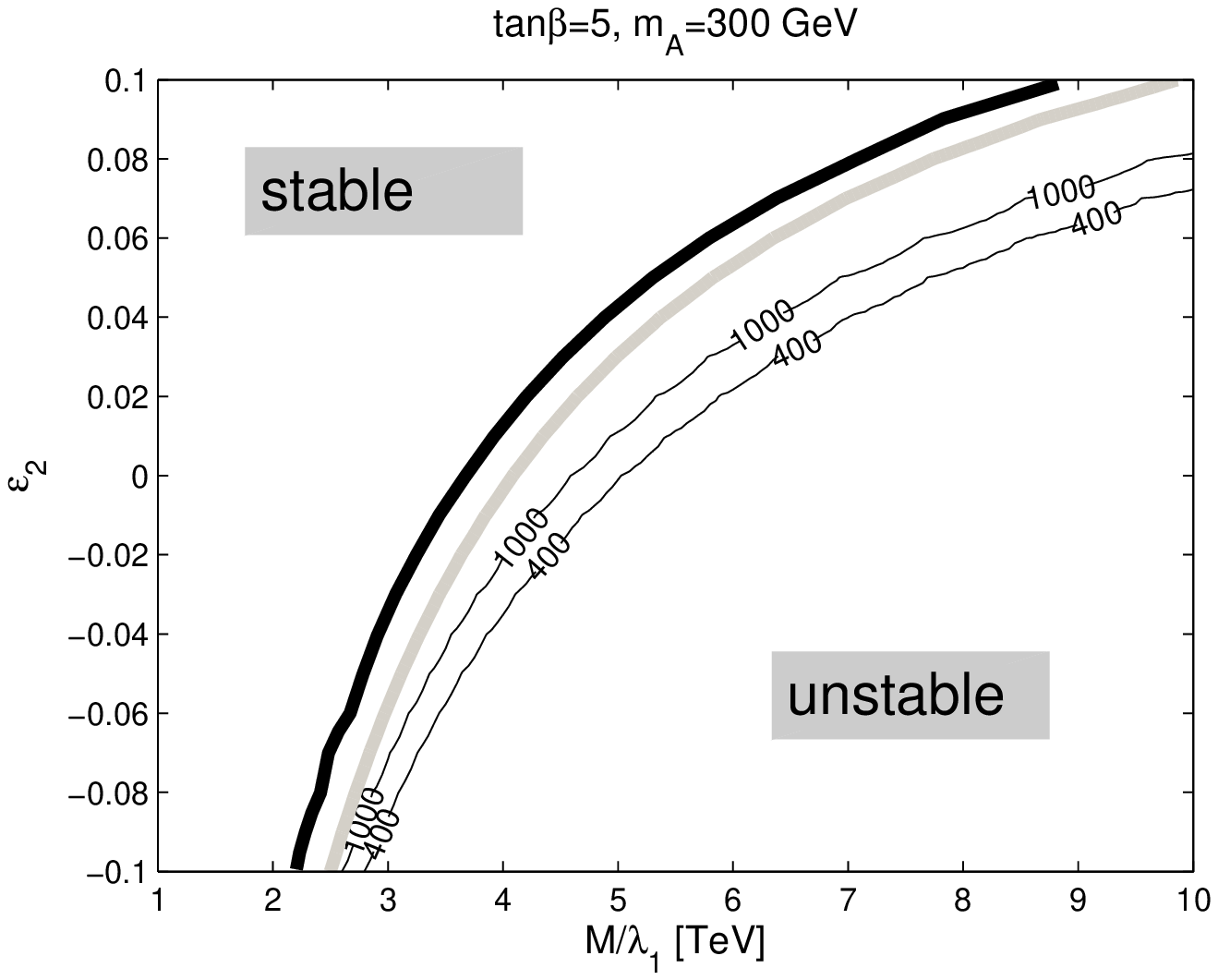}\hfill
\includegraphics[width=8.5cm]{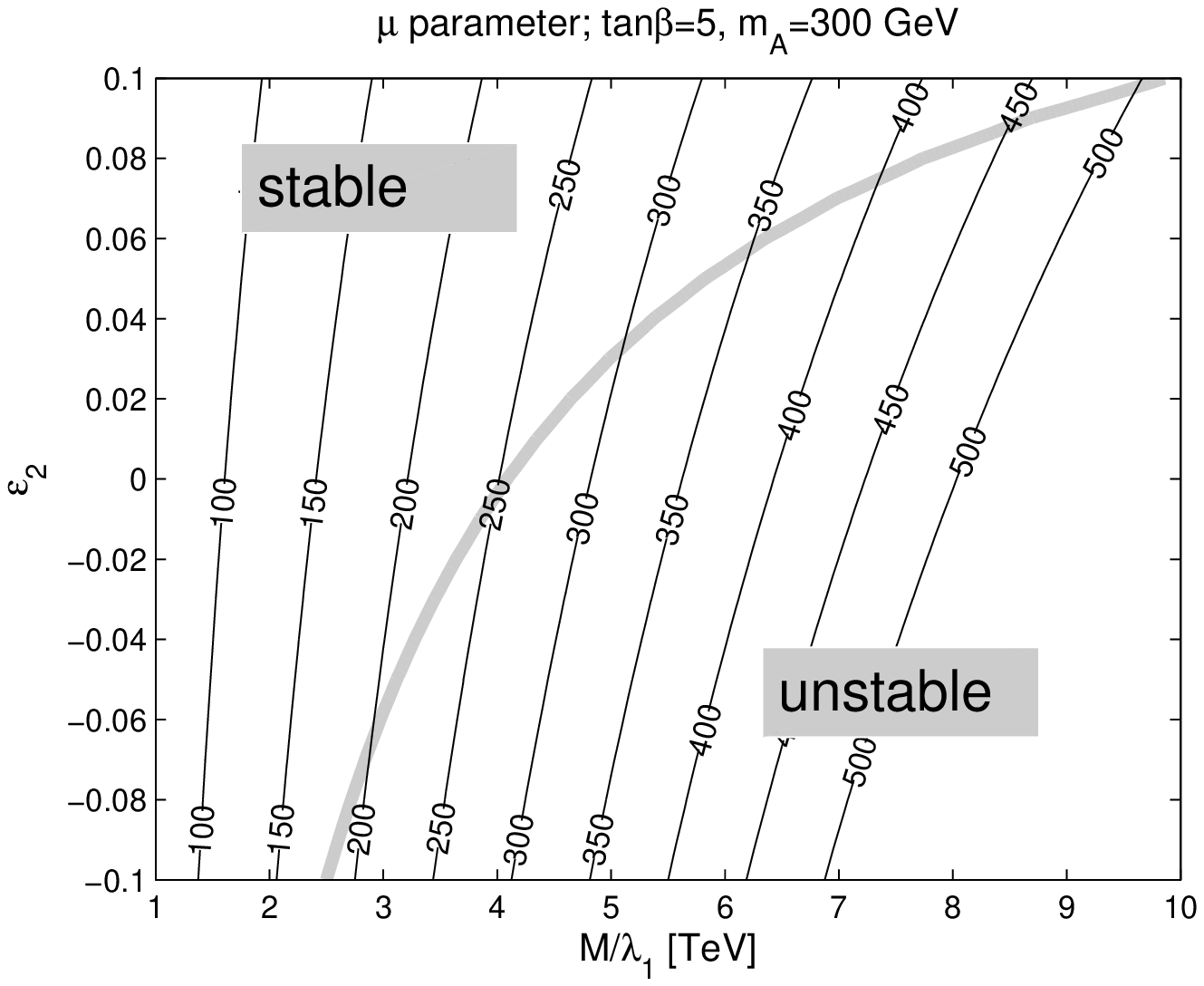}
\caption{Regions of vacuum stability, projected on the heavy scale $M$ vs. electroweak scale parameters.  All
plots are generated at tree level, for fixed value of $m_h=115$ GeV. Left panels:  Stability criteria of
Eq.~(\ref{eq:VstabEW}), vacuum degeneracy and quantum tunneling are denoted by thick gray, thick black and thin
labeled lines, respectively.  To the left of each line the EW vacuum is stable according to that criterion,
while to the right it is unstable.  Right panels: Contours of $|\mu|$ (in GeV) for the same parameter space of
the left panels. The stability criterion of Eq.~(\ref{eq:VstabEW}) is superimposed to enable easy comparison
between the two columns. }\label{fig:stabcritM}
\end{figure}

\section{Phenomenological implications}\label{sec:phen}

Finally the stage is set to study the implications of stability constraints. We begin by discussing relations
between electroweak scale parameters and the effective dimension five operators. We then study constraints on
the heavy BMSSM scale $M$. Lastly we describe the viable shift to the lightest Higgs mass, and outline how the
analysis is altered in the presence of radiative corrections.

\subsection{Constraints on BMSSM parameters}

The parameter space consistent with vacuum stability, depicted in Figure \ref{fig:stabcritM} for $m_h=115$ GeV,
becomes smaller when the value of $m_h$ is increased further above the experimental bound.\footnote{A quick way
to understand this statement (which will be assessed in Section \ref{ssec:mhvsM}) is by noting that, for given
values of $m_A,\tan\beta$ and $\epsilon_2$, the way to increase $m_h$ is via making $\epsilon_1$ more negative.
Vacuum instability is then driven stronger by the negative quartic $\sim\epsilon_1\phi^4$, balanced at large
fields by the dimension six term $\sim\phi^6/M^2$. It follows that the upper bound on $M$ derived from
stability must decrease as $m_h$ is increased.} Hence Figure \ref{fig:stabcritM} can be interpreted as a
translation of our current knowledge, $m_h>114$ GeV, into bounds on the viable ranges of other BMSSM
parameters, subject to the assumption that quantum corrections are small. Furthermore, one of the merits of the
analytical criterion Eq.~(\ref{eq:VstabEW}) is that it allows to extract limiting relations between BMSSM
parameters regardless of the precise value of the Higgs mass, as long as the expansion to order
$\mathcal{O}(\epsilon)$ is valid in the local vicinity of the EW vacuum. In order to gain further insight it is
useful to look at the behavior of Eq.~(\ref{eq:VstabEW}) in the following two limits.

\paragraph{Large $m_A$ limit -- $m_A^2\gg m_Z^2$:}
In this case Eq.~(\ref{eq:VstabEW}) implies
\beq\label{eq:mumA}|\mu|\lsim m_A\left|\frac{\epsilon_1}{\tilde\epsilon}\right|\sqrt{\frac{1+\sin2\beta}{2}}.
\eeq
This condition is robust but conservative. In other words, while having $|\mu|\lsim
m_A|\epsilon_1/\tilde\epsilon|$ ensures stability, somewhat larger values of $|\mu|/m_A$ may still yield stable
vacuum configurations, as illustrated in Figure \ref{fig:stabcritM}. Indeed, comparing the two upper panels we
see that while the vacuum degeneracy criterion requires $m_A>|\mu|$, the more loose tunneling action criterion
allows $m_A\approx|\mu|$. For example, taking $\tan\beta=4, \ \epsilon_2=-\epsilon_1=0.06,
\ |\mu|=m_A=3m_Z$ gives a stable vacuum configuration with $m_h\approx120$ GeV at tree level. Note that, in
order to guarantee vacuum stability, it is not a very good practice to assume very heavy $m_A$. The reason is
that in order for the effective theory to work properly, it is necessary to retain some hierarchy between the
high scale $M$ and electroweak mass parameters, and so taking any of the latter very large can pose a problem.
However, positive values for the SUSY breaking $\epsilon_2$ result in a cancellation in $\tilde\epsilon$ and so
the stability criteria can easily be satisfied through a combination of moderate ratios $m_A/|\mu|\sim1$ and
$|\epsilon_1/\tilde\epsilon|\gsim1$.

\paragraph{Small quartic coupling limit -- $|\tilde\epsilon/\epsilon_1|\ll1$:} An accidental cancellation between
the SUSY preserving and breaking dimension five operators can lead to $|\tilde\epsilon/\epsilon_1|\ll1$.
  In this case, for negative $\epsilon_1$, the RHS of Eq.~(\ref{eq:VstabEW}) approaches zero from above as the ratio
  $|\tilde\epsilon/\epsilon_1|$ decreases, and the stability constraint becomes
  trivially satisfied.
  A vanishing $\tilde\epsilon$ corresponds to large $\epsilon_2>0.1$
  if the LEP bound is satisfied via $\epsilon_1$. A finite ratio is then required for the effective expansion to
  remain valid. However, the role of the non-supersymmetric operator $\epsilon_2$ is clear. While it may not much affect
  the Higgs mass, this term may be important in stabilizing the potential; it thus partially decouples the stability
  problem from the spectrum.
  The importance of  $\epsilon_2$ for vacuum stability is illustrated in the lower panels of Figure \ref{fig:stabcritM}.
  We see that turning on even a small positive value for $\epsilon_2$ rapidly opens up the parameter space
  corresponding to a stable EW vacuum. Note also that the $\mu$ parameter is largely unaffected
  by a change of $\epsilon_2$ when the Higgs mass is held fixed. This reflects the fact that the stabilization of the
  potential occurs at large field values.

\subsection{Constraints on the scale of new physics}

We now proceed to discuss the constraints imposed by vacuum stability on the scale of new physics $M$.
To this end, note
that the $\mu$ parameter can be traded for $M$ using $\mu=\epsilon_1^*M/\lambda^*_1$. We find that imposing
stability implies an \emph{upper bound} on the heavy scale $M$. For example, assuming $m_A\gsim
3 m_Z$ we obtain from Eq.~(\ref{eq:VstabEW}):
\beq\label{eq:Mubnd}\left|\frac{M}{\lambda_1}\right|\lsim
\frac{m_A}{|\tilde\epsilon|}\sqrt{\frac{1+\sin2\beta}{2}}. \eeq
The upper bound on $M$ for various parameter settings can be read off of Figure \ref{fig:stabcritM}. Increasing
$m_A$, $\epsilon_2$ or lowering $\tan\beta$ all act to weaken the bound; this behavior is readily understood
from Eq.~(\ref{eq:Mubnd}). At $\tan\beta=5$ and with $|\tilde\epsilon|=0.05$, for instance, we obtain
$|M/\lambda_1|\lsim17m_A$.

Of course, while it is the main issue of the current paper, vacuum stability is not the sole source
of constraints on the scale of new physics. As an example, in Appendix \ref{app:EWPT} we show that electroweak
precision tests (EWPTs) result in a lower bound for the heavy scale $M$,
\beq\label{eq:EWPT} M\gsim 8  \ \mathrm{TeV.}\eeq
Put in conjunction with the stability constraints, a lower bound like (\ref{eq:EWPT}) points towards a large
value for $m_A$, sizable SUSY breaking $\epsilon_2$, a small $\tan\beta$ or some combination of the above.
Since the dimension six operators responsible for the leading Higgs mass shift and for potential stabilization
are different from the ones which affect EWPTs, the combined bounds on $M$ may be interpreted in two ways. On
the one hand, one may assume generic structure for the dimensionless coefficients of all nonrenormalizable
BMSSM operators. In this case, if quantum corrections are small ({\it e.g.}, say, both stops are found at the
LHC), then taking the bound (\ref{eq:EWPT}) together with the stability constraints leaves a very narrow range
for the scale of the heavy BMSSM physics. Conversely, one may interpret these two bounds as hinting towards
some suppression pattern in the microscopic extension behind the BMSSM theory.

\subsection{The lightest Higgs mass}\label{ssec:mhvsM}

The stability criteria can also be expressed in terms of constraints on the lightest Higgs mass. We choose to
present the resulting relation between $m_h$ and the heavy scale $M$. For clarity, we consider $m_A\gsim 3m_Z$
and expand to zeroth order in $m_Z^2/m_A^2$. Assuming a negative quartic $\tilde\epsilon<0$, we can convert
Eq.~(\ref{eq:Mubnd}) into a limit on $\epsilon_1$, put in terms of $m_A,\tan\beta,\epsilon_2$ and $M$. Plugging
the stability constraint in this form into the expression for the Higgs mass, Eq.~(\ref{eq:mhLargemA}), we
obtain
\beq\label{eq:mhvsM}
\frac{m_h^2}{m_Z^2}\lsim \cos^22\beta+\frac{4\epsilon_2\sin2\beta(1+\sin2\beta)}{g_Z^2}+
\frac{16m_A\sin2\beta\sqrt{(1+\sin2\beta)/2}}{g_Z^2|M/\lambda_1|}.
\eeq
It is illuminating to compare Eq.~(\ref{eq:mhvsM}) with Eq.~(\ref{eq:mhLargemA}). In particular, notice that in
the $\epsilon_2$ term a suppression factor of $\sin^22\beta$ appearing in the spectrum equation gets replaced
by a larger factor of $\sin2\beta(1+\sin2\beta)$ in the stability constraint. This result is  explained as
follows. First, recall that vacuum stability is trivially ensured if the overall quartic coupling along the
relevant D-flat direction is non-negative, $\tilde\epsilon\equiv\epsilon_2/4+\epsilon_1\geq0$. Given some value
of $\epsilon_2$, substituting $\epsilon_1=-\epsilon_2/4$ (which saturates the relation with a vanishing
$\tilde\epsilon$) into Eq.~(\ref{eq:mhLargemA}) gives the second term on the RHS of (\ref{eq:mhvsM}). Hence the
last term on the RHS of (\ref{eq:mhvsM}) represents the extra gain due to allowing a negative $\tilde\epsilon$
in the detailed analysis. This gain can indeed be significant: for example, at $\tan\beta=5(3)$ the numerical
coefficient in front of the ratio $m_A/|M/\lambda_1|$ equals $\sim9(16)$, so that a sizable shift for the Higgs
mass is possible even with $m_A\ll M$. In addition, if $\epsilon_2\approx0$ (in the case where the Beyond-MSSM
new physics threshold is supersymmetric, for instance) then the last term in Eq.~(\ref{eq:mhvsM}) provides the
maximal tree level mass shift consistent with vacuum stability.

\begin{figure}
\includegraphics[width=8.5cm]{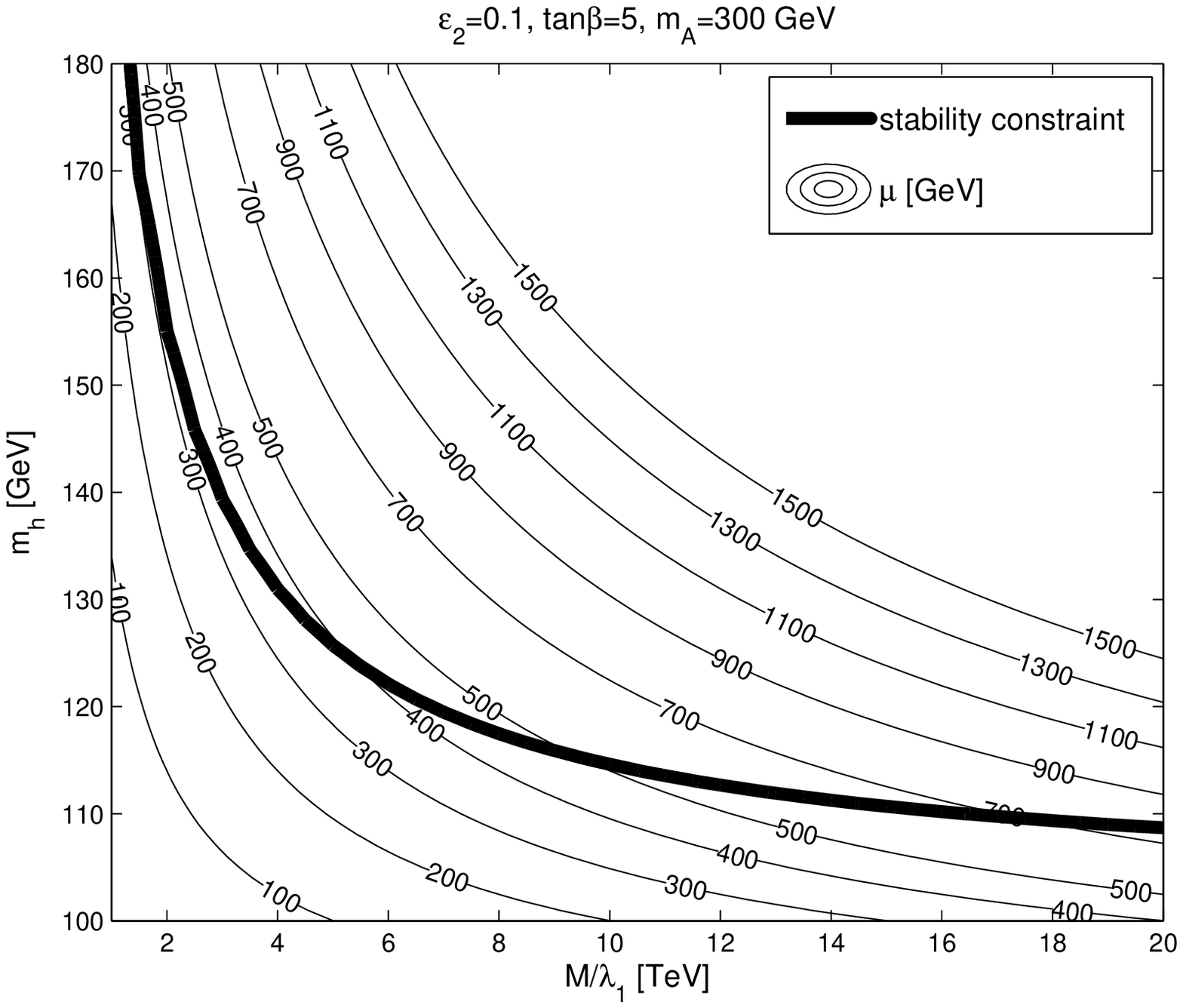}\hfill
\includegraphics[width=8.5cm]{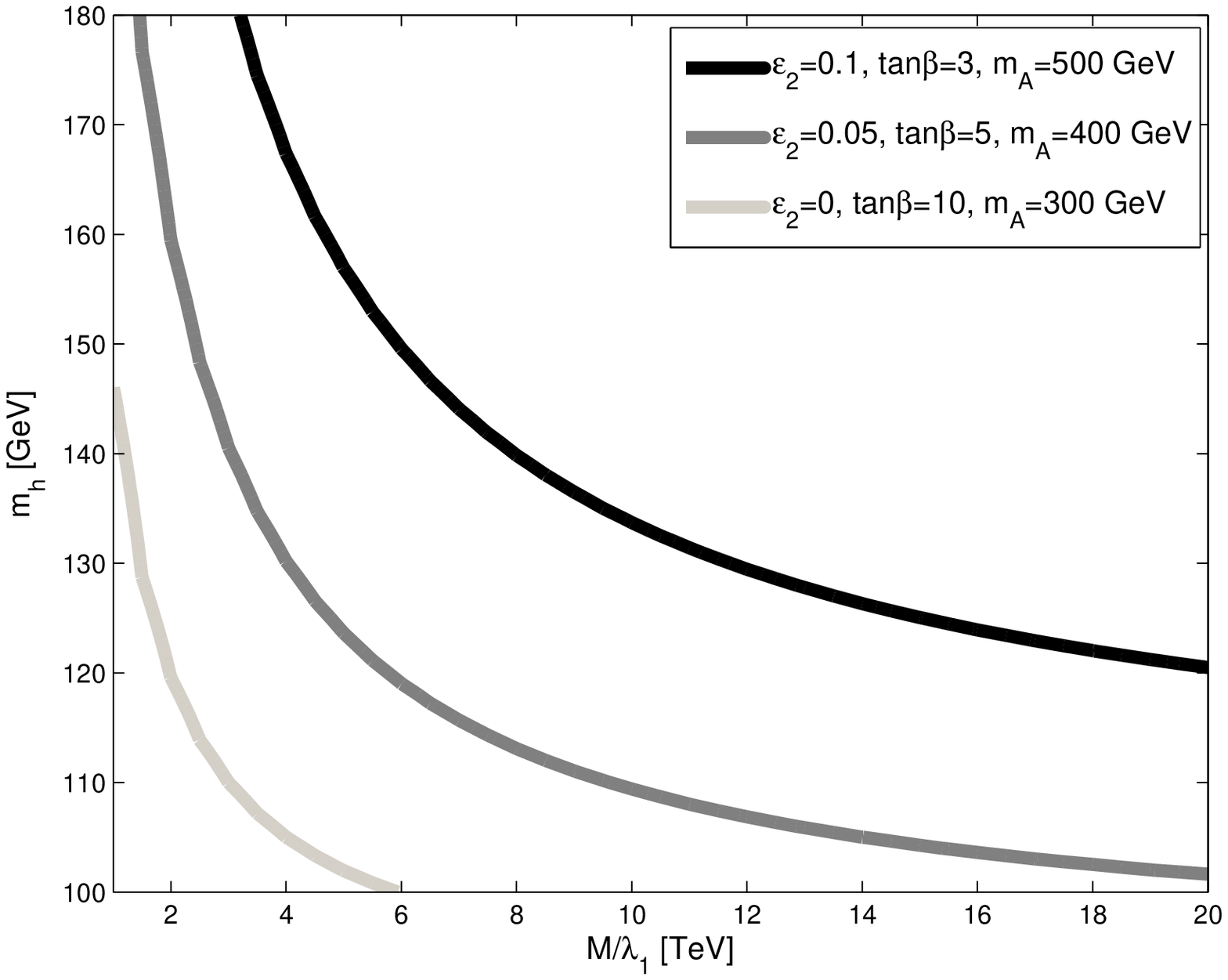}
\caption{Left: Stability constraint in the $(m_h,M)$ plane, for $m_A,\tan\beta$ and $\epsilon_2$ as indicated.
The thick line corresponds to the full analytical approximation (see text). Below this line the EW vacuum is
stable. Thin contours denote values of $|\mu|$. Right: Stability constraint for different values of
$\epsilon_2,m_A$ and $\tan\beta$.}\label{fig:mhvsM}
\end{figure}

The above analysis is illustrated in Figure~\ref{fig:mhvsM}. In the left panel we hold $m_A,\tan\beta$ and
$\epsilon_2$ fixed. The thick curve depicts the full analytical constraint derived from
Eqs.~(\ref{eq:mhLargemA}) and (\ref{eq:Mubnd}), for which Eq.~(\ref{eq:mhvsM}) represents the large $m_A$
limit. (It is easy to verify that Eq.~(\ref{eq:mhvsM}) follows this curve to a very good approximation.) In
order to keep track of the value of $\mu$, and so of $\epsilon_1$, we plot contours of constant $|\mu|$ as thin
labeled lines. In the right panel we explore the range of $m_h$ accessible in the BMSSM at tree level. We find
that with large $m_A$, small $\tan\beta$ and a sizable $\epsilon_2$, values of $m_h\approx140$ GeV can be
reached at tree level, in a stable vacuum configuration, even for $|M/\lambda_1|\gsim8$ TeV. Considering the
right panel of Figure~\ref{fig:mhvsM} and using Eq.~(\ref{eq:mhLargemA}), we find that $|\epsilon_1|$ becomes
larger than 0.1 at $m_h\gsim148,130,113$ GeV for the black, the dark gray and the light gray curves,
respectively. For larger values of $m_h$ the expansion to linear order in $\epsilon_1$ eventually begins to
break down. Thus even if very low values of $M/\lambda_1$ are acceptable, the stability curve cannot be trusted
to arbitrarily large $m_h$; we will touch upon this issue again in the next subsection.

\subsection{Quantum corrections}

\begin{figure}[h!]
\includegraphics[width=8.5cm]{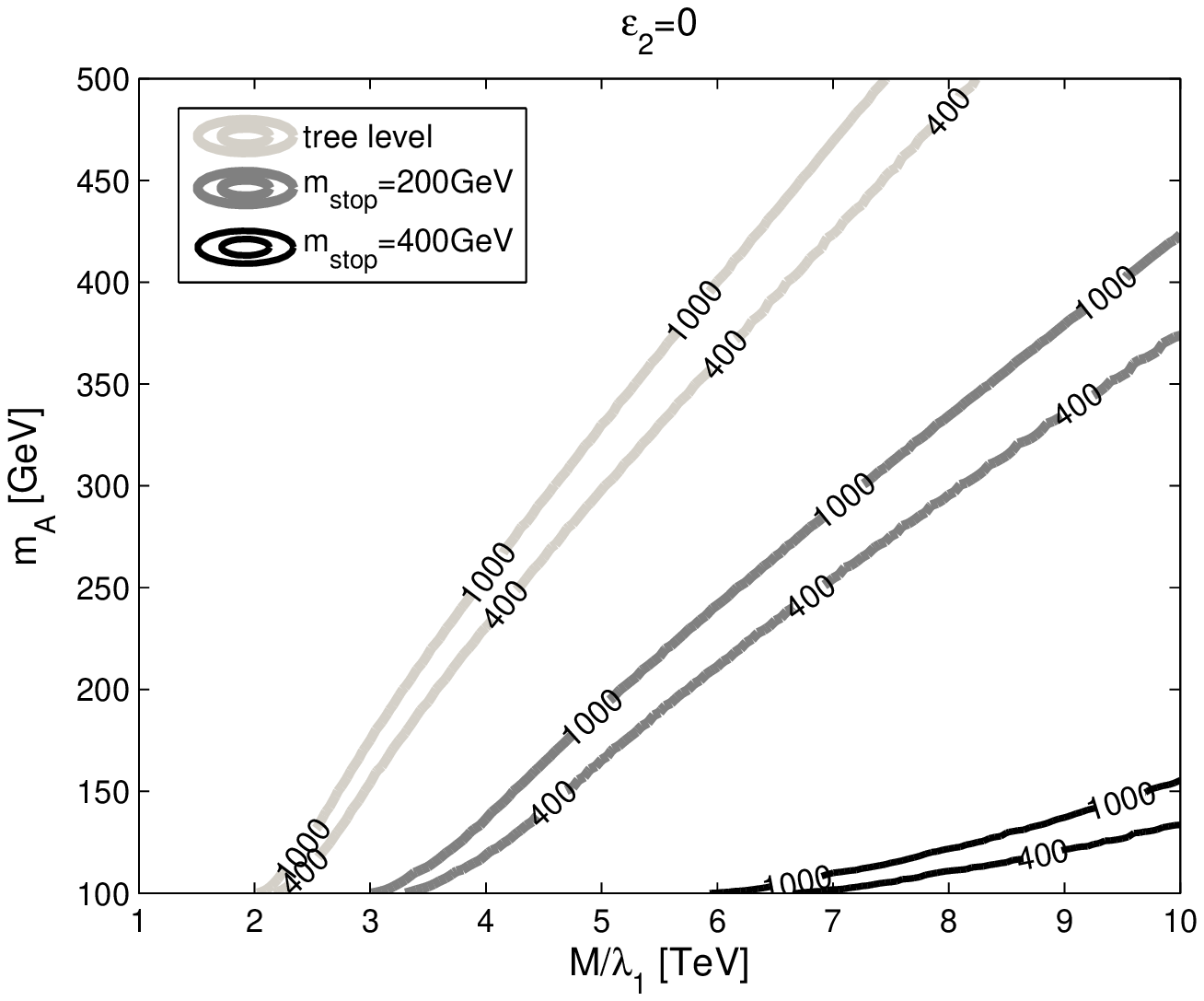}\hfill
\includegraphics[width=8.5cm]{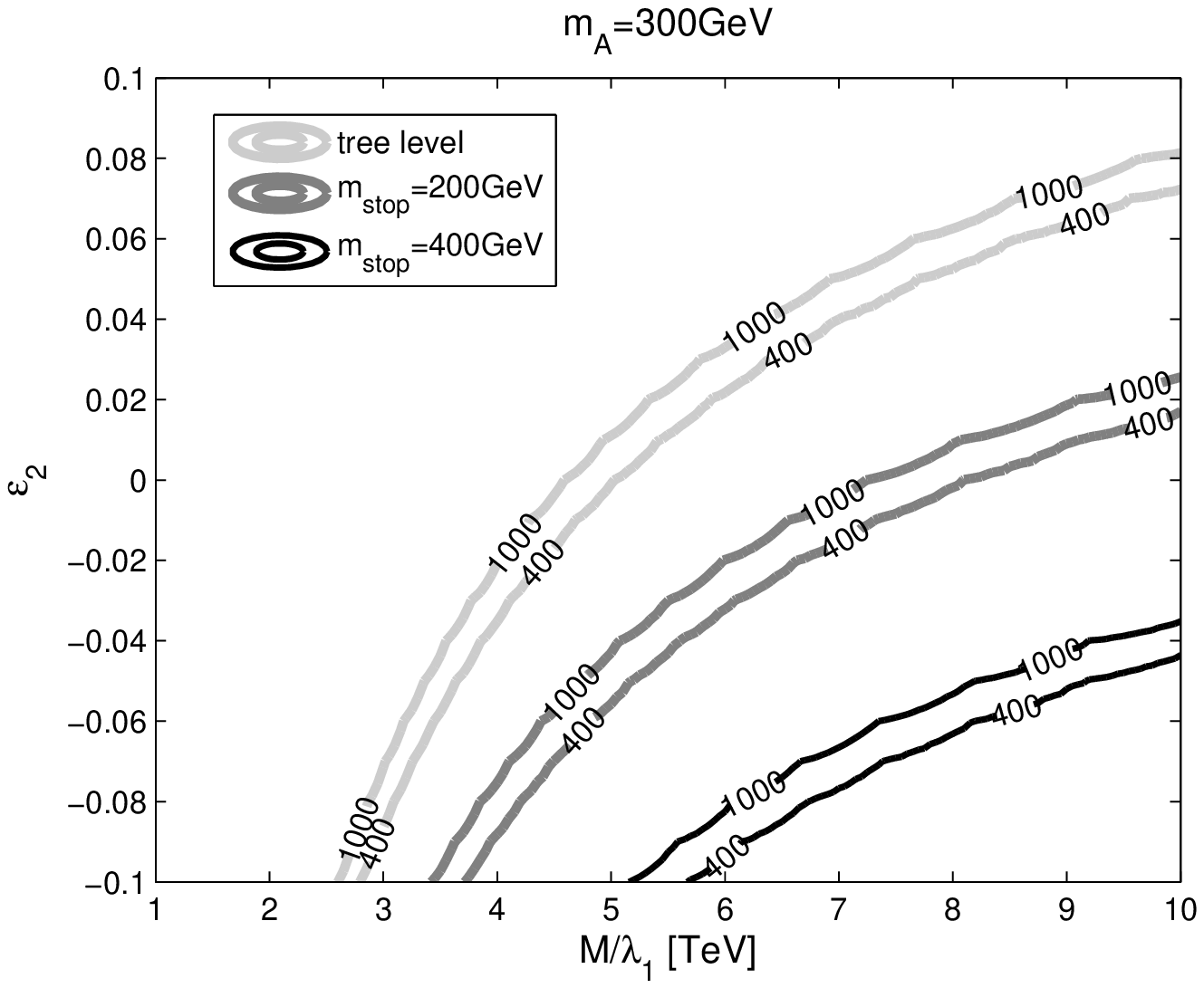}
\caption{Contours of the tunneling action, including quantum corrections due to top and stop loops
with a common soft stop mass of $0, \ 200 \mathrm{ \ and} \ 400$ GeV. The allowed parameter region lies
to the left of the contours. The plots are generated for $\tan\beta=5$ and $m_h=115$ GeV. }\label{fig:stabcritMQC}
\end{figure}

The radiative corrections to the scalar potential are dominated by top and stop loops in the moderate
$\tan\beta$ scenario. A (if not The) noteworthy feature of the BMSSM framework is that it allows quantum
corrections to be small \cite{Dine:2007xi}. As already implied in this work, a completely supersymmetric
top$-$stop sector is consistent with both the LEP bound on the lightest Higgs mass and stability
considerations. Nevertheless, it is interesting to check to what extent our results are affected by moderate
soft SUSY breaking masses for the stops. Here we do not attempt to study the stop sector in full detail. We
find it sufficient for our purpose to neglect stop mixing and D-terms, adopt a common soft mass for both stops
and work to one-loop order. Under these simplifications, quantum corrections modify the scalar potential by the
expression:
\beq\label{eq:QC}
\delta V_{\tilde{t}}\approx\frac{3}{16\pi^2}\left[m_{\tilde{t}}^4(\phi)\left(\ln\frac{m_{\tilde{t}}^2(\phi)}{Q^2}
-\frac{3}{2}\right)
-m_{t}^4(\phi)\left(\ln\frac{m_{t}^2(\phi)}{Q^2}-\frac{3}{2}\right)\right].
\eeq
The field dependent masses are given by $m_t(\phi)=y_t\phi_2, \
m^2_{\tilde{t}}(\phi)=m_t^2(\phi)+m^2_{\mathrm{stop}}$, with $m_{\mathrm{stop}}$ the soft stop mass and $y_t$
the top Yukawa. $Q$ is the renormalization scale, which we choose to be $m_Z$. In writing Eq.~(\ref{eq:QC}), we
impose renormalization conditions such that Eq.~(\ref{eq:vtb}) remains unchanged. In Figure
\ref{fig:stabcritMQC} we repeat the stability analysis with quantum corrections, focusing for concreteness on
the $(M,m_A)$ and $(M,\epsilon_2)$ planes. We learn that soft masses for stops stabilize the potential,
effectively removing the tree level upper bound on $M$ for $m_{\tilde{t}}\gsim 400$ GeV.

\begin{figure}[h!]\begin{center}
\includegraphics[width=8.5cm]{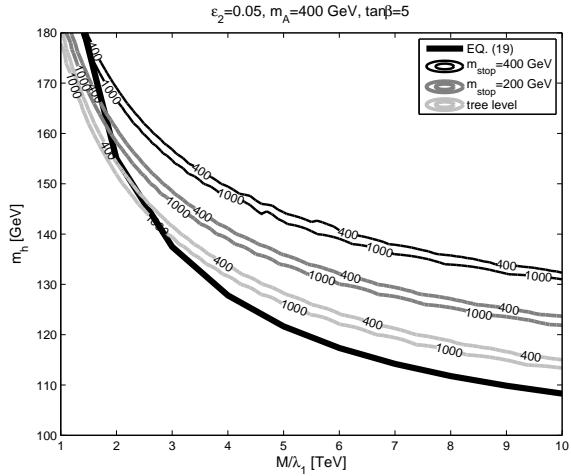}
\caption{Contours of the tunneling action, including quantum corrections due to top and stop loops with a
common soft stop mass of $0, \ 200 \mathrm{ \ and} \ 400$ GeV. The allowed parameter region lies below the
contours. The plot is generated for fixed $m_A,\tan\beta$ and $\epsilon_2$ as indicated.
}\label{fig:stabcritMhMQC} \end{center}\end{figure}

Finally, Figure \ref{fig:stabcritMhMQC} depicts the effect of quantum corrections on the results derived for
the lightest Higgs boson mass. Comparing with the right panel of Figure \ref{fig:mhvsM} we find, as expected,
that quantum corrections increase the upper bound on $m_h$. A combination of small BMSSM operators
$\epsilon\lsim0.05$ with modest radiative corrections from stops at the 200$-$250 GeV range can easily and
naturally accommodate the experimental bound on $m_h$, even for $M\gsim10$ TeV. Figure~\ref{fig:stabcritMhMQC}
includes the tree level numerical stability criterion of the tunneling action, as well as the analytical
criterion Eq.~(\ref{eq:mhvsM}). Comparing the two we find that the latter provides a reasonable, though
slightly conservative approximation to the numerical bound for values of $m_h$ smaller than $\sim140$ GeV. For
$m_h\gsim140$ GeV, $|\epsilon_1|$ becomes larger than $\sim0.13$ and the expansion to  $\mathcal{O}(\epsilon)$
begins to fail, demonstrating the regime of validity of our analytical arguments.

%

\section{Conclusions}\label{sec:disc}

In this paper we analyzed the vacuum structure of the BMSSM Higgs sector. We showed that the effective
dimension five operators which lift the lightest Higgs mass are potentially harmful, as they are capable of
destabilizing the scalar potential. It is easy to ensure that MSSM D-terms prevent the instability from
occurring over most of the field space, for scales $\phi<M$, by imposing $\epsilon^2\lsim1/15$. This condition
does not exclude a significant shift to the Higgs mass. Furthermore, it stands in accordance with the desire to
keep the effective theory expansion under control. Along the D-flat directions, however, the MSSM D-terms
vanish and the leading BMSSM correction is dominant. Thus if the quartic coupling along one of these directions
is negative the effective expansion must be taken beyond leading order. Scrutinizing the effective theory to
order $1/M^2$, we were able to show that the stability of the scalar potential is controlled by only one
scaling dimension six operator, which is supersymmetric and positive definite.  This operator is correlated
with the supersymmetric effective dimension five term which is the most relevant for lifting the lightest Higgs
mass. Hence the nonrenormalizable part of the theory can cure itself, even though a remote vacuum may emerge
before the cutoff scale of the effective theory.

In order to deal with a non-trivial potential structure, a set of criteria was developed from which relations
between BMSSM parameters were inferred guaranteeing vacuum stability. In particular, by analyzing the potential
along the D-flat directions we derived an approximate analytical criterion, Eq.~(\ref{eq:VstabEW}), whose
robustness was demonstrated by means of a full numerical study. Using this criterion we showed, for example,
that if the LEP bound on the Higgs mass is accommodated at tree level in the BMSSM, then the stability of the
EW vacuum is ensured provided that $m_A\gtrsim|\mu|$. Additionally, very low values of
$\tan\beta\sim\mathcal{O}(1)$ are allowed and even favored despite the diminished MSSM contribution to the
Higgs mass. Interestingly, at the classical level, vacuum stability implies an \emph{upper bound} on the heavy
scale $M$. This bound is better defined than what abstract fine-tuning arguments would suggest, and, since the
experimental data can be accommodated with small $\mu/M\lsim0.05$, is also typically stronger. Put in
conjunction with generic lower bounds on $M$, arising for instance from electroweak precision tests, the
analysis either highly constrains the BMSSM parameter space or directs us towards a non-generic coupling
structure in the effective theory.

Vacuum stability also dictates an upper bound on the lightest Higgs mass accessible via the leading effective
operators. However, for $M$ in the few TeV range, $m_h\gsim140$ GeV can still be accommodated at tree level, as
a result of the potential stabilization provided by the supersymmetric dimension six operator. While a
completely supersymmetric top$-$stop sector is allowed in the BMSSM (as far as the experimental bound on the
lightest Higgs mass is regarded), quantum corrections due to stop loops are effective in stabilizing the
potential, and two stop states at $\sim300$ GeV suffice to significantly weaken the constraints due to vacuum
stability. Indeed, light stops just around the corner for upcoming colliders can provide a relatively
unimportant direct contribution for the Higgs spectrum, yet be significant for vacuum stabilization at large
field values.

Finding both stops not too far above $m_t$ would already be a smoking gun for a Beyond-MSSM extension.
Information on the spectrum of charginos, neutralinos and the Higgs scalars, together with stability
considerations, could then be used in order to extract additional constraints on dimension five operators and
low energy MSSM parameters, as well as put an upper bound on the heavy scale $M$.

\section*{Acknowledgments}

We thank Yossi Nir for fruitful discussions and Michael Dine, Zohar Komargodski and Nathan Seiberg for useful comments on the
manuscript.

\appendix

\section{CP violating and charge breaking background fields}\label{app:AppPotential}

Under the assumption of real parameters, the full effective potential including $\chi$ and $\rho$ fields is
given by
\beq\label{eq:Vfull}
V(\phi_1,\phi_2,\chi,\rho)&=&m_1^2(\phi_1^2+\chi^2+\rho^2)
+m_2^2\phi_2^2-2\left[m_{12}^2+2\epsilon_1\left(\phi_1^2+\chi^2+\rho^2+\phi_2^2\right)\right]\phi_1\phi_2 \nn\\
&&+2\epsilon_2\left(\phi_1^2-\chi^2\right)\phi_2^2
+\frac{g_Z^2}{8}\left(\phi_1^2+\chi^2+\rho^2-\phi_2^2\right)^2+\frac{g^2}{2}\phi_2^2\rho^2\nn\\
&&+4\left|\frac{\epsilon_1}{\mu}
\right|^2\phi_2^2\left(\phi_1^2+\chi^2+\rho^2+\phi_2^2\right)\left(\phi_1^2+\chi^2\right).
\eeq
First derivatives of interest follow,
\beq\label{eq:DVchi2}
  \frac{\partial V}{\partial {\chi^2}}&=&m_1^2-4\epsilon_1\phi_1\phi_2-2\epsilon_2\phi_2^2
  +\frac{g_Z^2}{4}\left(\phi_1^2+\chi^2+\rho^2-\phi_2^2\right)\nn\\ &&
  +4\left|\frac{\epsilon_1}{\mu}\right|^2\phi_2^2\left(2\phi_1^2+2\chi^2+\phi_2^2+\rho^2\right),
\eeq
\beq\label{eq:DVrho2}
  \frac{\partial V}{\partial {\rho^2}}&=&m_1^2-4\epsilon_1\phi_1\phi_2+\frac{g_Z^2}{4}\left(\phi_1^2+\chi^2+
  \rho^2-\phi_2^2\right)
  +4\left|\frac{\epsilon_1}{\mu}\right|^2\phi_2^2\left(\phi_1^2+\chi^2\right)+\frac{g^2}{2}\phi_2^2.
\eeq
Since only the squares of $\chi$ and $\rho$ appear in the potential, a trivial extremum solution with
$\chi=\rho=0$ always exists. CP violating (CPV) and charge breaking (CB) extrema require nontrivial solutions
for a vanishing RHS in Eqs.~(\ref{eq:DVchi2}) and (\ref{eq:DVrho2}), respectively.

We first derive the condition ensuring that the EW vacuum is CP and charge conserving. Demanding that the
potential has a minimum at $\phi_1=v\cos\beta, \ \phi_2=v\sin\beta$ and $\rho=\chi=0$ and expanding to leading
order in $\epsilon$, we obtain that an additional, nontrivial solution with $\phi_1=v\cos\beta, \
\phi_2=v\sin\beta$ and $\chi\neq0$ is never allowed, while a solution with $\rho\neq0$ requires
$\rho^2\simeq-4\sin^2\beta\left(m^2_A+m_W^2+2\epsilon_2v^2\right)/g_Z^2$. Hence $\chi=\rho=0$ is the only
consistent solution at the EW vacuum, as long as: \beq m^2_A+m_W^2+2\epsilon_2v^2>0.\eeq Considering
Eq.~(\ref{eq:sc4eff6}) and the validity of the effective theory expansion, we find that this relation is always
satisfied for $m_A\gtrsim m_Z$.

We now move on to find out how nonzero $\chi$ or/and $\rho$ affect the potential at large field values. In
particular, we are interested in configurations of $\chi,\rho$ which extremize the potential in the vicinity of
the tunneling path. Finding such extrema is important, since they may shift the saddle point and/or the remote
vacuum from the $(\phi_1,\phi_2)$ plane and so alter the numerical computation of the stability criteria. Note
that for large fields, unlike in the local neighborhood of the EW vacuum, one must keep the scaling dimension
six operator appearing in the potential (\ref{eq:Vfull}). In addition to the trivial solution $\{\chi= 0,\rho=
0\}$, there are three exclusive possibilities: $\{\chi\neq 0,\rho= 0\}$, $\{\chi=0,\rho\neq 0\}$ or $\{\chi\neq
0,\rho\neq 0\}$. To find out which of them is consistent, we rewrite the condition of vanishing right hand
sides in Eqs.~(\ref{eq:DVchi2}) and (\ref{eq:DVrho2}) as
\beq\chi^2=A_{CPV}(\phi_1,\phi_2)+B_{CPV}(\phi_1,\phi_2)\rho^2,\quad\rho^2=A_{CB}(\phi_1,\phi_2)+B_{CB}(\phi_1,\phi_2)
\chi^2.\eeq
Solutions with $\{\chi\neq 0,\rho= 0\}$ or $\{\chi=0,\rho\neq 0\}$ only exist for $A_{CPV}>0$ or $A_{CB}>0$
respectively. Moreover, since the second derivatives of the potential with respect to $\chi^2$ and $\rho^2$ are
positive definite, whenever one of these solutions becomes available within some region of the
$(\phi_1,\phi_2)$ plane it will always minimize the potential energy, rendering the trivial solution
unfavorable. Hence if the tunneling path traverses these regions, CPV or CB will turn on. A solution with
$\{\chi\neq 0,\rho\neq 0\}$ requires that both $\chi^2=(A_{CPV}-A_{CB}B_{CPV})/(1-B_{CB}B_{CPV})$ and
$\rho^2=(A_{CB}-A_{CPV}B_{CB})/(1-B_{CB}B_{CPV})$ be positive. If such a solution exists, it may or may not
become energetically favored over solutions with only $\chi$ or $\rho$ nonzero.

We now describe how the $\chi$ and $\rho$ fields are dealt with in the numerical procedures presented in
Section \ref{ssec:numer}. The criterion of vacuum degeneracy requires computing the value of the potential at
the remote minimum. For the tunneling action, according to the approximation we adopt (see Appendix
\ref{app:qtunn}), the location of and potential value at both the remote minimum and saddle point are needed.
To find these extrema in the presence of possible CPV and CB configurations, we numerically evaluate the full
potential over the $(\phi_1,\phi_2)$ plane where $\chi$ and $\rho$ are replaced by the four classes of
solutions defined above. Then we retain all consistent solutions extremizing the potential along both the
$\phi_1$ and $\phi_2$ directions. While the remote minimum is unique, as it has to minimize the potential along
all directions, there may be several saddle points corresponding to the different classes of solutions for
$\chi$ and $\rho$. We compute the tunneling action assuming the path passes through the saddle point of minimum
potential energy.

After the dust settles it turns out that in typical cases in which the EW vacuum is found to be stable, no CPV
or CB arise in regions of field space that are relevant for the analysis, or otherwise the effect is very
small. Thus the discussion conducted in Section \ref{ssec:anal}, where for clarity we have set $\chi=\rho=0$,
holds true up to minor changes. In order to see how this result comes about, it is useful once again to
consider the problem close to the D-flat directions. Since away from them positive D-terms raise the potential,
the tunneling path is confined to remain near this region. Thus removing CPV and CB from the flat directions
protects the saddle point and the remote vacuum from developing nonzero $\chi$ or $\rho$. We can derive simple
analytical criteria to exclude the formation of CPV or CB extrema along the D-flat directions. To this end, let
us consider $\epsilon_1<0$ in which case potential instability occurs for negative $\phi_1$. Setting
$-\phi_1=\phi_2\equiv\phi/\sqrt{2}$ and $\tilde\epsilon\equiv\epsilon_2/4+\epsilon_1<0$, the extremum equations
reduce to:
\beq
-\left(\frac{g_Z^2}{4}+4\left|\frac{\epsilon_1}{\mu}\right|^2\phi^2\right)\chi^2=m_1^2+4\left(\frac{3\epsilon_1}{2}-\tilde{\epsilon}\right)\phi^2+3\left|\frac{\epsilon_1}{\mu}\right|^2\phi^4
+\left(\frac{g_Z^2}{4}+2\left|\frac{\epsilon_1}{\mu}\right|^2\phi^2\right)\rho^2,
\eeq
\beq
-\frac{g_Z^2}{4}\rho^2=m_1^2+\frac{g^2}{4}\left(1+\frac{8\epsilon_1}{g^2}\right)\phi^2+\left|\frac{\epsilon_1}{\mu}\right|^2\phi^4+\left(\frac{g_Z^2}{4}+2\left|\frac{\epsilon_1}{\mu}\right|^2\phi^2\right)\chi^2.
\eeq We find that CPV and CB are banned along the D-flat directions provided the following conditions hold,
respectively:
\beq\label{eq:NoCPV} \frac{m_A^2(1+\sin2\beta)}{|\mu|^2} \geq
\frac{4}{3}f(\beta)\left(\frac{\tilde{\epsilon}}{\epsilon_1}\right)^2\left(1-
\frac{3\epsilon_1}{2\tilde\epsilon}\right)^2,\quad \frac{m_A^2(1+\sin2\beta)}{|\mu|^2} \geq
f(\beta)\left(\frac{g^2}{8\epsilon_1}+1\right)^2 \eeq
up to $\mathcal{O}(m_Z^2/m_A^2)$ corrections, where $f(\beta)\equiv(1+\sin2\beta)/\sin^2\beta$ satisfies $1<f(\beta)<4$ for $\tan\beta>1$. 
These two conditions can be compared to the approximate analytical criterion derived in Section \ref{ssec:anal}
which ensures that the EW vacuum is stable, namely
$m_A^2(1+\sin2\beta)/|\mu|^2\geq2(\tilde\epsilon/\epsilon_1)^2$.  Regarding CB, we find that vacuum stability
guarantees no CB turns on along the D-flat direction provided that $|\epsilon_1|\lsim 0.2$.  In view of Figure
\ref{fig:epscontours}, as well as considering Eq.~(\ref{eq:sc4eff6}) and the validity of the effective theory
expansion, it is clear that this relation does not constitute a real compromise for the size of $\epsilon_1$.
The appearance of CPV is a less marginal effect.  However, the form of the inequalities (\ref{eq:NoCPV}) and
the stability constraint implies that the stable parameter space exhibits little CPV; numerically we find that
in most stable scenarios CPV does not occur along the tunneling path, or else does not encompass the saddle
point nor the remote minimum. We conclude that imposing vacuum stability typically renders CB and CPV
configurations irrelevant in the calculation. Note that since the tunneling path does not pass strictly along,
but merely in the vicinity of the D-flat direction, the above analysis provides an intuitive argument only.
However, as previously mentioned, numerically scanning the relevant parameter space we find that in practice it
is indeed safe to neglect $\chi$ and $\rho$ in the analysis of vacuum stability in the BMSSM with real
Lagrangian parameters.

\section{Quantum tunneling}\label{app:qtunn}

We are interested in analyzing the stability of configurations wherein the potential exhibits a remote vacuum,
in the presence of which our EW vacuum may be metastable. At zero temperature, one should compute the rate of
quantum tunneling from the EW to the remote vacuum. For completeness, let us briefly review the theoretical set up
before going into the details of our implementation.
In a semi-classical approach the tunneling rate (per unit time and volume) is given by the WKB
approximation~\cite{Coleman:1977py}:
\beq\label{eqn:Gammarate} \Gamma\simeq A\exp\left(-B\right), \ \ \ B=S_E[\phi_b]-S_E[\phi_{\rm false}], \eeq
where $S_E[\phi]$ is the Euclidean action
and $A$ is the determinant of the Gaussian fluctuations around $\phi_b$, the so-called bounce solution of the
equation of motion. The precise value of the prefactor $A$ plays a minor role in comparison to the exponential
suppression factor. Hence we make the conservative assumption that the prefactor saturates the highest scale in
the BMSSM framework: $A\simeq M^4$. The bounce solution is $O(4)$ symmetric~\cite{Coleman:1977th}, {\it i.e.}
is only a function of $r=\sqrt{x^2+t_E^2}$. It solves the equation of motion
\beq \frac{d^2\phi_b}{dr^2}+\frac{3}{r}\frac{d\phi_b}{dr}-\nabla V(\phi_b)=0, \eeq
subject
to the ``bouncing'' boundary conditions: $\phi'_b(0)=0$ and $\phi_b(\infty)=\phi_{\rm false}$.

In configuration space, quantum tunneling proceeds by nucleation, growth and percolation of true vacuum bubbles
surrounded by a metastable environment. At a given time $t$, the portion $\mathcal{P}(t)$ of the volume of the
universe filled by true vacuum bubbles is controlled by the tunneling rate and the expansion of the universe
through the following relation~\cite{Guth:1981uk}
\beq \mathcal{P}(t)= 1-\exp\left[-\frac{4\pi}{3}\int_0^t dt'\Gamma(t')d_H^3(t')\right], \eeq
where $d_H(t)\sim t$ is the horizon
distance. For the case of quantum tunneling ($T=0$), the rate $\Gamma$, given by (\ref{eqn:Gammarate}), is time
independent. Hence today, for $t_0\sim 10^{10}{\rm yr}\sim e^{100} v^{-1}$, the portion of the universe in
the stable phase is given approximately by:
\beq \mathcal{P}(t_0)\sim
1-\exp\left[-\mathcal{O}(10)\exp\left(4\log\frac{M}{v}-B\right)(vt_0)^4\right]. \eeq
Therefore to ensure that
the metastable vacuum has a life time longer than the present age of the universe, or equivalently that
$\mathcal{P}(t_0)\simeq 0$, it is enough to impose the following constraint on the bounce action: \beq B\gtrsim
400+4\log\frac{M}{v}, \eeq where the second term, coming from the exponential prefactor $A$,
contributes at most $\mathcal{O}(10\%)$ for $M$ in the TeV range.

In a one-dimensional field space (as in the Standard Model), the correct bounce solution is easily found by use of a
simple ``shooting'' numerical method. However, in the case of multiple fields, the search of the bounce
solution is a non-trivial numerical task. We make use of several simplifying assumptions, justified in the
BMSSM setup under consideration.

\begin{figure}[h!tb]
\includegraphics[scale=0.7]{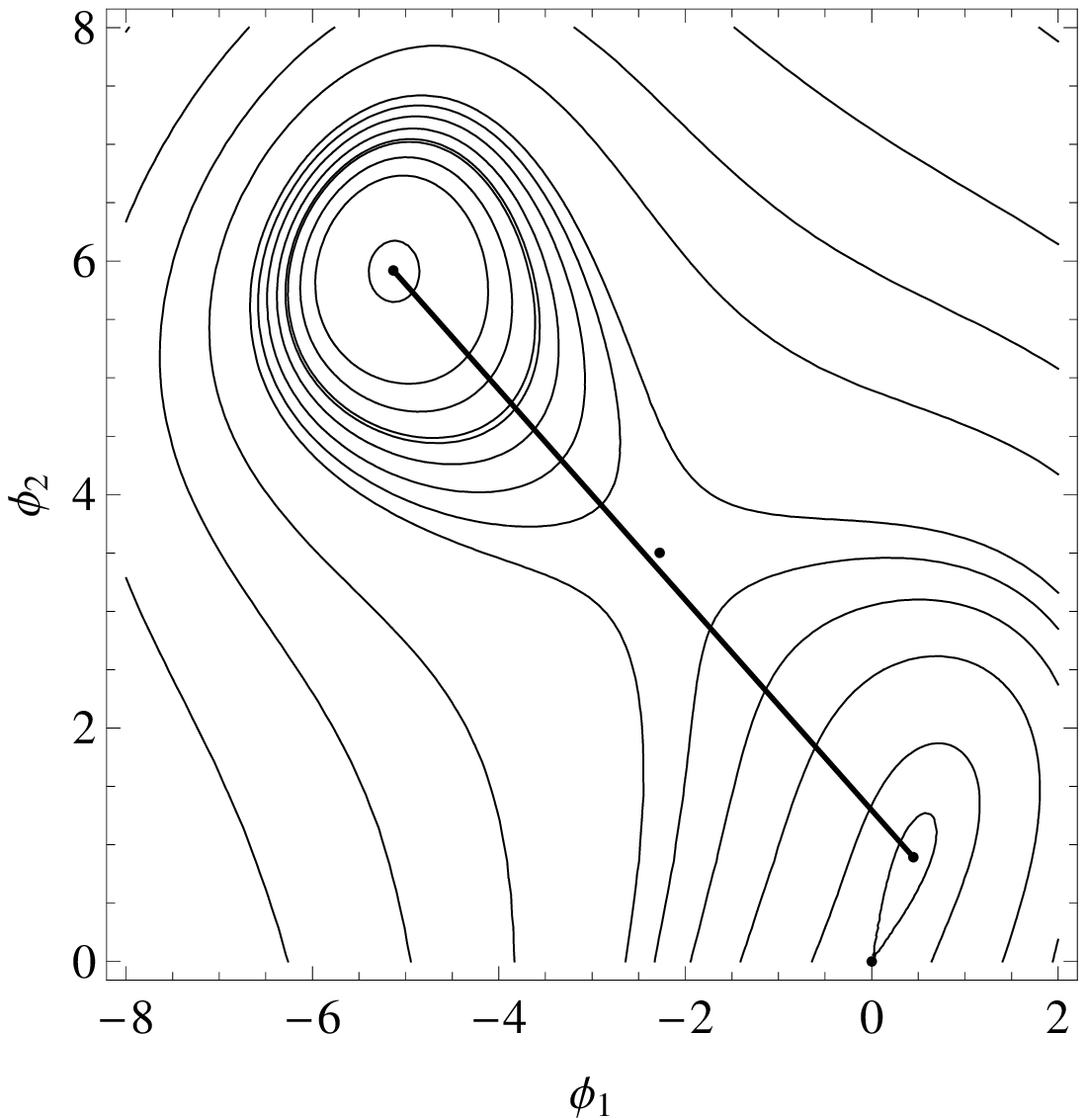}\hfill
\includegraphics[scale=0.7]{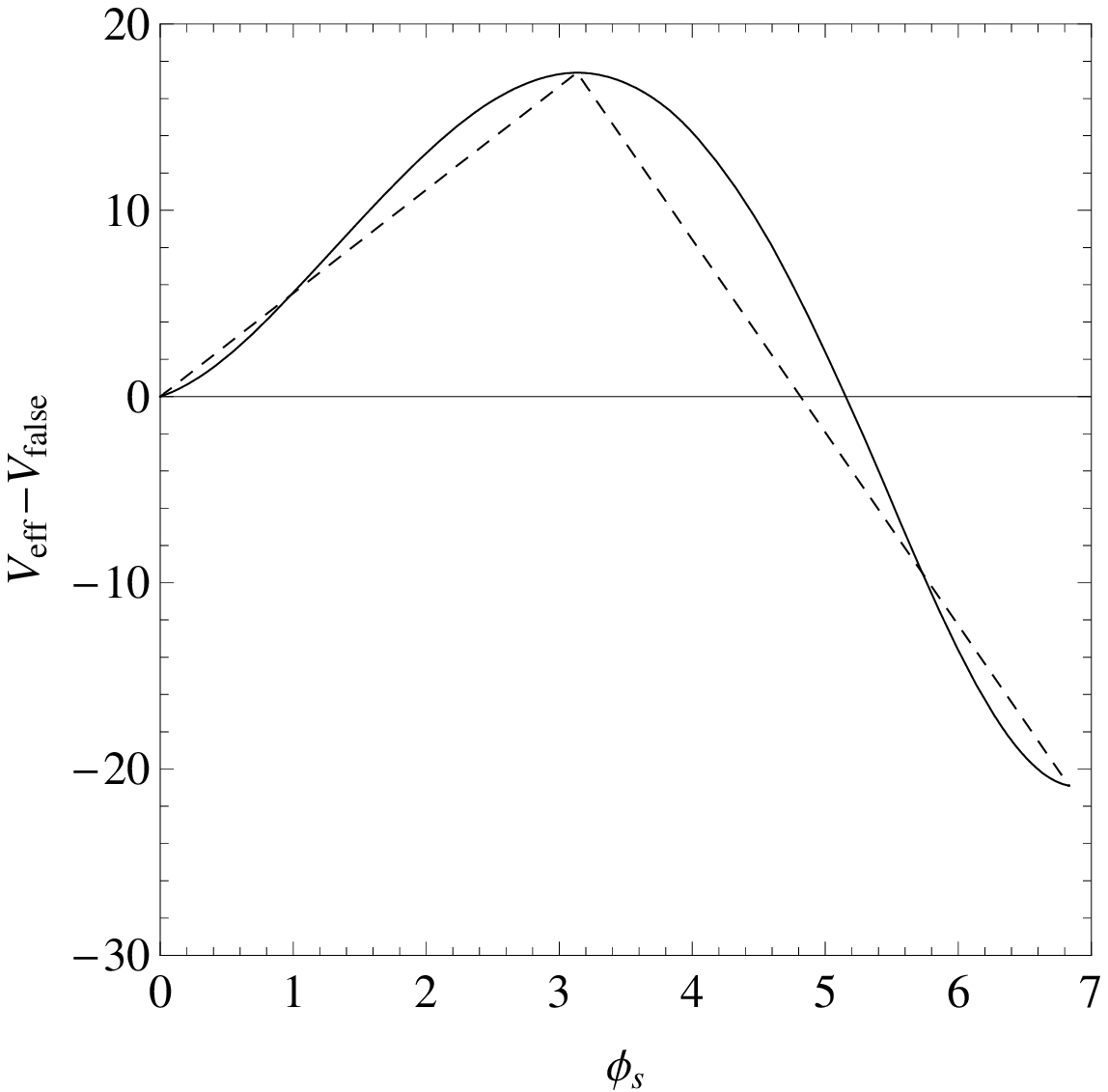}
\caption{Left: Contour plot of the potential for $\mu=m_A=300$ GeV, $\tan\beta=2$, $\epsilon_2=0$ and
$\epsilon_1\simeq-0.05$, corresponding to tree level Higgs mass of $m_h\simeq 115$ GeV. The straight line
connects the two minima. Right: Potential projected on the straight line joining the two vacua (solid) and the
corresponding triangle approximation (dashed) used to estimate the tunneling action. $\phi_s$ is the coordinate
along the straight line. On both plots $\phi$ and $V_{\rm eff}$ are in units of $v$ and $v^4$
respectively.}\label{fig:Vplots}
\end{figure}

The first simplification consists of projecting the action on a given path in field space, which is known to be
close to the correct tunneling trajectory. This path is fully characterized by a single curvilinear coordinate
$\phi_s$, {\it i.e.} $\phi=\phi(\phi_s)$. The initial task is then greatly simplified as the problem has been
reduced to a one dimensional search for the bounce solution. However such a projection is in general hard to
estimate. One exception is when the potential extrema are close to alignment in field space. In this case the
tunneling trajectory may be well approximated by a straight line joining the true and false
vacua~\cite{Banks:1974ij}. Projecting on such a path allows one to easily compute the bounce action using a
traditional shooting method. We stress that this approach is consistent as the approximate alignment that it
requires typically arises in the BMSSM. Indeed, as is apparent in Figure~\ref{fig:Vplots}, both the remote vacuum and
the saddle point separating it from the EW vacuum form close to the D-flat direction.

Second, we adopt a triangular approximation for the potential barrier, allowing us to use the clear prescription given
in~\cite{Duncan:1992ai}. Thus we reserve the more cumbersome shooting method only as a check on the triangle
calculation, verifying that the two methods agree at the level of $\mathcal{O}(10\%)$. In Figure \ref{fig:Vplots}
we demonstrate the use of the triangle approximation in computing the tunneling action.
%

\section{Electroweak constraints}\label{app:EWPT}

As we have seen, stability of the scalar potential, arising from
effective dimension six operators, gives an upper bound on the scale
of new physics. Other effective dimension six operators involving
the Higgs sector also affect the gauge terms in the
Lagrangian, introducing mass shifts in the gauge sector as well as
kinetic mixings. They are therefore constrained by electroweak data,
in particular the precision electroweak variables $S$ and $T$
\cite{Amsler:2008zz} \beq S=-0.10\pm
0.10\ \ ,\ \ T=-0.08\pm 0.11 \eeq
for $m_h\simeq 115$ GeV.
There are several effective dimension six operators involving the Higgs sector that can contribute to
electroweak observables.  For example, operators of the form $a_{ij}(H_i^\dagger e^V H_i)(H_j^\dagger e^V
H_j)/M^2$ with $i,j=u,d$ in the $\mathrm{K\ddot{a}hler}$ potential contribute to the deviation of the $\rho$
parameter from unity by \cite{Brignole:2003cm}
\beq\label{eq:rho}
\rho-1=-4\left(\frac{m_W}{gM}\right)^2\left[a_{uu}\sin^4\beta+a_{dd}\cos^4\beta-a_{ud}\cos^2\beta\sin^2\beta\right]\simeq \alpha T.
\eeq
For generic coefficients $a_{ij}\sim {\cal O}(1)$, the expression in
square brackets in \eqref{eq:rho} is not particularly suppressed for
any specific value of $\tan\beta$.  We obtain a lower
bound on the heavy scale,
\beq\label{eq:EWPTbound} M\gsim 7.7  \ \mathrm{TeV.}\eeq

Regarding the $S$ parameter, consider for instance a superpotential operator of the form $a_{WB}W^\alpha B_\alpha H_u
H_d/M^2$, where $W_\alpha(B_\alpha)$ denotes the $SU(2)_L$($U(1)_Y$) gauge superfield
strength. Such an operator generates a contribution to $S$ of \beq S=\frac{32\pi}{g g'}\frac{v^2
\sin2\beta}{2M^2}a_{WB}. \eeq Taking $a_{WB}\sim {\cal O}(1)$ we obtain a comparable bound, $M\gsim 8\sqrt{\sin2\beta}$
TeV.


\end{document}